%% file: main.tex
\documentclass[runningheads]{llncs}
\usepackage[paperheight=235mm, paperwidth=155mm,textwidth=12.2cm,textheight=19.3cm]{geometry}
\usepackage{enumitem}
\usepackage{amsmath}
\usepackage{amssymb}
\usepackage{mathtools}
\usepackage[table,dvipsnames]{xcolor}
\usepackage{kbordermatrix}
\usepackage{tikz}
\usetikzlibrary{arrows.meta}
\usetikzlibrary{arrows,automata}
\usetikzlibrary{shapes,decorations,patterns,positioning}
\usepackage{subcaption}
\usepackage{multirow}
\usepackage{pgfplots}
\usepackage{todonotes}
\pgfplotsset{compat=1.13}
\renewcommand{\kbldelim}{(}
\renewcommand{\kbrdelim}{)}

\makeatletter
\RequirePackage[bookmarks,unicode,colorlinks=true]{hyperref}%
   \def\@citecolor{blue}%
   \def\@urlcolor{blue}%
   \def\@linkcolor{blue}%

\def\orcidID#1{\smash{\href{http://orcid.org/#1}{\protect\raisebox{-1.25pt}{\protect\includegraphics{ORCID_Color.eps}}}}}
\makeatother

\usepackage{bbding}

\def\techreport{}
\title{Correlated Equilibria and Fairness in Concurrent Stochastic Games}

\ifthenelse{\isundefined{\techreport}}{\author{Marta Kwiatkowska\inst{1}\orcidID{0000-0001-9022-7599} \and Gethin Norman\inst{1,2}\orcidID{0000-0001-9326-4344} \and David~Parker\inst{3}\orcidID{0000-0003-4137-8862} \and Gabriel~Santos\inst{1}(\Envelope)\orcidID{0000-0002-6570-9737}}
}{
\author{Marta Kwiatkowska\inst{1} \and Gethin Norman\inst{1,2} \and David~Parker\inst{3} \and Gabriel Santos\inst{1}}
}
\authorrunning{Marta Kwiatkowska \and Gethin Norman \and David~Parker \and Gabriel Santos}

\institute{Department of Computer Science, University of Oxford, Oxford, UK \email{\{marta.kwiatkowska,gabriel.santos\}@cs.ox.ac.uk}
\and
School of Computing Science, University of Glasgow, Glasgow, UK
\email{gethin.norman@glasgow.ac.uk}
\and School of Computer Science, University of Birmingham, Birmingham, UK
\email{d.a.parker@cs.bham.ac.uk}}

\input{macros-dave.tex}

\begin{document}

\maketitle

\begin{abstract}
Game-theoretic techniques and equilibria analysis facilitate the design and verification of competitive systems.
While algorithmic complexity of equilibria computation has been extensively studied, practical implementation and application of game-theoretic methods is more recent.
Tools such as PRISM-games support automated verification and synthesis of zero-sum and ($\varepsilon$-optimal subgame-perfect) social welfare Nash equilibria properties for concurrent stochastic games.
However, these methods become inefficient as the number of agents grows
and may also generate equilibria that yield significant variations in the outcomes for individual agents.
We extend the functionality of PRISM-games to support \emph{correlated equilibria}, in which players can coordinate through public signals, and introduce a novel optimality criterion of \emph{social fairness}, which can be applied to both Nash and correlated equilibria.
We show that correlated equilibria are easier to compute,
are more equitable, and can also improve joint outcomes.
We implement algorithms for both normal form games
and the more complex case of multi-player concurrent stochastic games with temporal logic specifications.
On a range of case studies, we demonstrate the benefits of our methods.
\end{abstract}

\section{Introduction}\label{sec:intro}

Game-theoretic verification techniques can support the
modelling and design of systems that comprise multiple
agents operating in either a cooperative or competitive manner.
In many cases, to effectively analyse these systems we also
need to adopt a probabilistic approach to modelling,
for example because agents operate in uncertain environments,
use faulty hardware or unreliable communication mechanisms,
or explicitly employ randomisation for coordination.

In these cases, \emph{probabilistic model checking} provides
a convenient unified framework for both
formally modelling probabilistic multi-agent systems
and 
specifying their required behaviour.
In recent years, progress has been made in this direction
for several models, including turn-based and concurrent stochastic games (TSGs and CSGs),
and for multiple temporal logics, such as rPATL~\cite{CFK+13b} and its extensions~\cite{KNPS21}.
Tool support has been developed, in the form of PRISM-games~\cite{KNPS20},
and successfully applied to case studies across a broad range of areas.

Initially, the focus was on \emph{zero-sum} specifications~\cite{KNPS21},
which can be natural for systems whose participants have
directly opposing goals, such as the defender and attacker in
a security protocol 
minimising or maximising the probability of a successful attack, respectively.
However, agents often have objectives that are distinct
but not directly opposing,
and may also want to cooperate to achieve these objectives.
Examples include network protocols and multi-robot systems.

For these purposes, \emph{Nash equilibria} (NE) have also been integrated into
probabilistic model checking of CSGs~\cite{KNPS21},
together with \emph{social welfare} (SW) optimality criterion, resulting in 
social welfare Nash equilibria (SWNE).
An SWNE comprises a strategy for each player in the game
where no player has an incentive to deviate unilaterally from their strategy
and the sum of the individual objectives over all players is maximised.

One key limitation of SWNE, however, is that, as these techniques are extended
to support larger numbers of players~\cite{KNPS20b}, the efficiency and scalability of synthesising
SWNE is significantly reduced.
In addition, simply aiming to maximise the sum of individual objectives
may not produce the best performing equilibrium,
either collectively or individually;
for example, they can offer higher gains for specific players,
reducing the incentive of the other players to collaborate and instead motivating them to deviate from the equilibrium.

In this paper, we adopt a different approach and introduce, for the first time within formal verification, both \emph{social fairness} as an optimality criterion and \emph{correlated equilibria}, and the insights required to make these usable in practical applications. Social fairness (SF) is particularly novel, as it is inspired by similar concepts used in economics and distinct from the fairness notions employed in verification.
Correlated equilibria (CE)~\cite{Aum74},
in which players are able to coordinate through public \emph{signals}, are easier to compute than NE and can yield better outcomes.
Social fairness,
which minimises the differences
between the objectives of individual players, can be considered for both CE and NE.

We first investigate these concepts for the simpler case of
normal form games, illustrating their differences and benefits. We then extend the approach to the more
powerful modelling formalism of CSGs and extend the
temporal logic rPATL to formally specify agent objectives.
We present algorithms to synthesise equilibria,
using linear programming to find CE
and a combination of backwards induction or value iteration for CSGs.
We implement our approach in the PRISM-games tool~\cite{KNPS20}
and demonstrate significant gains in computation time and that quantifiably more fair and useful strategies can by synthesised for a range of application domains.
This paper is an extended version of \cite{confversion}, including the complete model checking algorithm.

\startpara{Related work}
Nash equilibria have been considered for concurrent systems in~\cite{GHW14}, where a temporal logic is proposed whose 
key operator 
is a novel path quantifier which asserts that a property holds on all Nash equilibrium computations of the system. There is no stochasticity and correlated equilibria are not considered. 
In \cite{AKMMR19}, a probabilistic logic that can express equilibria is formulated, along with complexity results, but no implementation has been provided.

The notion of fairness studied here is
inspired by fairness of equilibria from economics~\cite{Rab93,Rab97}  
and aims to minimise the difference between the payoffs, as opposed to maximising the lowest payoff among the players in an NE~\cite{LRTZ06}.
Our notion of fairness can be thought of as a constraint applied to equilibria strategies, similar in style to social 
welfare, and used to select certain equilibria based on optimality. 
This is distinct from fairness 
used in 
verification of concurrent processes, where
(strong) fairness refers to a property stating that, whenever a process is enabled infinitely often, it is executed infinitely often. This notion is typically defined as a constraint on infinite execution paths expressible in logics LTL and CTL* and needed to prove liveness properties. 
For probabilistic models, verification under fairness constraints has been formulated for Markov decision processes and the logic PCTL* \cite{BK98,BK08}. %
For games on graphs, fairness conditions expressed as $\omega$-regular winning conditions can be used to synthesise reactive processes  \cite{CF11}.
Algorithms for strong transition fairness for $\omega$-regular games have been recently studied in 
\cite{BMMSS21}.
Both qualitative and quantitative approaches have been considered for verification under fairness constraints, but no equilibria.

\section{Normal Form Games}\label{nfg-sect}

We start by considering normal form games (NFGs), then define our equilibria concepts for these games,
present algorithms and an implementation for computing them,
and finally summarise some experimental results.

We first require the following notation. Let $\dist(X)$ denote the set of probability distributions over set $X$. For any vector $v \in \Rset^n$, we use $v(i)$ to refer to the $i$th entry of the vector. For any tuple $x=(x_1,\dots,x_n) \in X^n$, element $x' \in X$ and $i \leq n$, we define the tuples $x_{-i} \rmdef (x_1,\dots,x_{i-1},x_{i+1},\dots,x_n)$ and $x_{-i}[x'] \rmdef (x_1,\dots,x_{i-1},x',x_{i+1},\dots,x_n)$.

\begin{definition}[Normal form game] 
A (finite, $n$-person) \emph{normal form game} (NFG) is a tuple $\nfgame = (N,A,u)$ where: $N=\{1,\dots,n\}$ is a finite set of players; $A = A_1 {\times} \cdots {\times} A_n$ and $A_i$ is a finite set of actions available to player $i \in N$; $u = (u_1,\dots,u_n)$ and $u_i \colon A \rightarrow \Rset$ is a utility function for player $i \in N$.
\end{definition}
We fix an NFG $\nfgame= (N,A,u)$ for the remainder of this section. 
In a play of $\nfgame$, each player $i \in N$ chooses an action from the set $A_i$ at the same time. If each player $i$ chooses $a_i$, then the utility received by player $j$ equals $u_j(a_1,\dots,a_n)$.
We next define the \emph{strategies} for players of $\nfgame$ and \emph{strategy profiles} comprising a strategy for each player. We also define \emph{correlated profiles}, which allow the players to coordinate their choices through a (probabilistic) \emph{public signal}. 
\begin{definition}[Strategy and profile]\label{strats-nfgs}
A \emph{strategy} $\sigma_i$ for player $i$ is an element of $\Sigma_i = \dist(A_i)$ and a \emph{strategy profile} $\sigma$ is an element of $\Sigma_\nfgame = \Sigma_1{\times}\cdots{\times}\Sigma_n$.
\end{definition}
For strategy $\sigma_i$ of player $i$, the \emph{support} is the set of actions $\{ a_i \in A_i \mid \sigma_i(a_i){>}0 \}$ and the support of a profile is the product of the supports of the strategies.
\begin{definition}[Correlated profile]\label{corp-nfgs}
A \emph{correlated profile} is a tuple $(\tau,\varsigma)$ comprising $\tau \in \dist(D)$, where $D=D_1{\times}\cdots{\times}D_n$, $D_i$ is a finite set of \emph{signals} for player $i$,
and $\varsigma=(\varsigma_1,\dots,\varsigma_n)$, where $\varsigma_i \colon D_i \ra A_i$.
\end{definition}
For a correlated profile $(\tau,\varsigma)$, the public signal $\tau$ is a joint distribution over signals $D_i$ for each player $i$ such that, if player $i$ receives the signal $d_i \in D_i$, then it chooses action $\varsigma_i(d_i)$. 
We can consider any correlated profile $(\tau,\varsigma)$ as a \emph{joint strategy}, i.e., a distribution over $A_1 {\times}\cdots{\times}A_n$ where:
\[ \begin{array}{c}
(\tau,\varsigma)(a_1,\dots,a_n) = \sum \{ \tau(d_1,\dots,d_n) \mid d_i \in D_i \wedge \varsigma(d_i)=a_i \; \mbox{for all $i \in N$} \} \, .
\end{array}
\]
Conversely, any joint strategy $\tau \in \dist(A_1 {\times}\cdots{\times}A_n)$ can be considered as a correlated profile $(\tau,\varsigma)$ where $D_i=A_i$ and $\varsigma_i$ is the identity function for $i \in N$.

Any strategy profile $\sigma$ can be mapped to an equivalent correlated profile
(in which $\tau$ is the joint distribution $\sigma_1 {\times}\cdots {\times} \sigma_n$ and $\varsigma_i$ is the identity function). On the other hand, there are correlated profiles with no equivalent strategy profile. Under profile $\sigma$ and correlated profile $(\tau,\varsigma)$ the expected utilities of player $i$ are:
\[ 
\begin{array}{rcl}
u_i(\sigma) & \rmdef & \sum_{(a_1,\dots,a_n) \in A} u_i(a_1,\dots,a_n) \cdot \big( \prod_{j=1}^n \sigma_j(a_j) \big) \\
u_i(\tau,\varsigma) & \rmdef & \sum_{(d_1,\dots,d_n) \in D} \tau(d_1,\dots,d_n) \cdot u_i(\varsigma_1(d_1),\dots,\varsigma_n(d_n)) \, .
\end{array} 
\]
\begin{examp}
Consider the two-player NFG where $A_i = \{ a_1^i , a_2^i \}$ and a correlated profile corresponding to the joint distribution $\tau \in \dist(A_1{\times}A_2)$ where $\tau(a_1^1,a_2^1) = \tau(a_1^2,a_2^2) = 0.5$. 
Under this correlated profile the players share a fair coin and both choose their first action if the coin is heads and their second action otherwise. This has no equivalent strategy profile.
\end{examp}
\startpara{Optimal equilibria of NFGs} We now introduce the notions of \emph{Nash equilibrium}~\cite{NMK+44} and \emph{correlated equilibrium}~\cite{Aum74}, as well as different definitions of optimality for these equilibria: \emph{social welfare} and \emph{social fairness}. Using the notation introduced above for tuples, for any profile $\sigma$ and strategy $\sigma^\star_i$, the strategy tuple $\sigma_{-i}$ corresponds to $\sigma$ with the strategy of player $i$ removed and $\sigma_{-i}[\sigma^\star_i]$ to the profile $\sigma$ after replacing player $i$'s strategy with $\sigma^\star_i$.
\begin{definition}[Best response]\label{def:bestresponse} 
For a profile $\sigma$ and correlated profile $(\tau,\varsigma)$, 
a \emph{best response} for player $i$ to $\sigma_{-i}$ and $(\tau,\varsigma_{-i})$ are, respectively:
\begin{itemize}
\item
a strategy $\sigma^\star_i$ for player $i$ such that $u_i(\sigma_{-i}[\sigma^\star_i]) \geq u_i(\sigma_{-i}[\sigma_i])$ for all $\sigma_i \in \Sigma_i$;
\item
a function $\varsigma^\star_i \colon D_i \ra A_i$ for player $i$ such that $u_i(\tau,\varsigma_{-i}[\varsigma^\star_i]) \geq u_i(\tau,\varsigma_{-i}[\varsigma_i])$ for all functions $\varsigma_i \colon D_i \ra A_i$.
\end{itemize}
\end{definition}
\begin{definition}[NE and CE]\label{def:eq}
A strategy profile $\sigma^\star$ is a \emph{Nash equilibrium} (NE) and a correlated profile $(\tau,\varsigma^\star)$ is a \emph{correlated equilibrium} (CE) if:
\begin{itemize}
\item
$\sigma_i^\star$ is a best response to $\sigma_{-i}^\star$ for all $i \in N$;
\item
$\varsigma_i^\star$ is a best response to $(\tau,\varsigma_{-i}^\star)$ for all $i \in N$;
\end{itemize}
respectively. We denote by $\Sigma^N$ and $\Sigma^C$ the set of NE and CE, respectively.
\end{definition}
Any NE of $\nfgame$ is also a CE, while there can exist CEs that cannot be represented by a strategy profile and therefore are not NEs. 
For each class of equilibria, NE and CE, we introduce two optimality criteria, the first maximising \emph{social welfare} (SW), defined as the \emph{sum} of the utilities, and the second maximising \emph{social fairness} (SF), which minimises the \emph{difference} between the players' utilities. Other variants of fairness have been considered for NE, such as in \cite{LRTZ06}, where the authors seek to maximise the lowest utility among the players.
\begin{definition}[SW and SF]\label{def:swne}
An equilibrium $\sigma^\star$ is a \emph{social welfare} (SW) equilibrium if the sum of the utilities of the players under $\sigma^\star$ is maximal over all equilibria, while $\sigma^\star$ is a \emph{social fair} (SF) equilibrium if the difference between the player's utilities under $\sigma^\star$ is minimised over all equilibria.
\end{definition}
We can also define the dual concept of \emph{cost equilibria}~\cite{KNPS21}, where players try to minimise, rather than maximise, their expected utilities by considering equilibria of the game $\nfgame^{-}= (N,A,{-}u)$ in which the utilities of $\nfgame$ are negated.
\begin{figure}[t]
\begin{subfigure}{0.49\textwidth}
\centering
\includegraphics{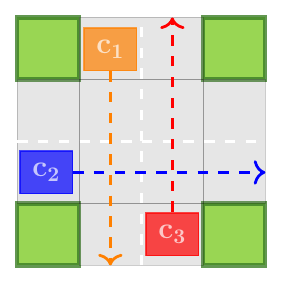}
\end{subfigure}
\centering
\begin{subfigure}{0.49\textwidth}
{\scriptsize
\begin{tabular}{cccc}
$a$                                 & $u_1(a)$                  & $u_2(a)$                  & $u_3(a)$                  \\ \hline
\multicolumn{1}{|c|}{$({\textit{pro}}_1, {\textit{pro}}_2, {\textit{pro}}_3)$} & \multicolumn{1}{c|}{$-1000$} & \multicolumn{1}{c|}{$-1000$} & \multicolumn{1}{c|}{$-100$} \\ \hline
\multicolumn{1}{|c|}{$({\textit{pro}}_1, {\textit{pro}}_2, {\textit{yld}}_3)$} & \multicolumn{1}{c|}{$-1000$} & \multicolumn{1}{c|}{$-100$} & \multicolumn{1}{c|}{$-5$} \\ \hline
\multicolumn{1}{|c|}{$({\textit{pro}}_1, {\textit{yld}}_2, {\textit{pro}}_3)$} & \multicolumn{1}{c|}{$5$} & \multicolumn{1}{c|}{$-5$} & \multicolumn{1}{c|}{5} \\ \hline
\multicolumn{1}{|c|}{$({\textit{pro}}_1, {\textit{yld}}_2, {\textit{yld}}_3)$} & \multicolumn{1}{c|}{$5$} & \multicolumn{1}{c|}{$-5$} & \multicolumn{1}{c|}{$-5$} \\ \hline
\multicolumn{1}{|c|}{$({\textit{yld}}_1, {\textit{pro}}_2, {\textit{pro}}_3)$} & \multicolumn{1}{c|}{$-5$} & \multicolumn{1}{c|}{$-1000$} & \multicolumn{1}{c|}{$-100$} \\ \hline
\multicolumn{1}{|c|}{$({\textit{yld}}_1, {\textit{pro}}_2, {\textit{yld}}_3)$} & \multicolumn{1}{c|}{$-5$} & \multicolumn{1}{c|}{$5$} & \multicolumn{1}{c|}{$-5$} \\ \hline
\multicolumn{1}{|c|}{$({\textit{yld}}_1, {\textit{yld}}_2, {\textit{pro}}_3)$} & \multicolumn{1}{c|}{$-5$} & \multicolumn{1}{c|}{$-5$} & \multicolumn{1}{c|}{$5$} \\ \hline
\multicolumn{1}{|c|}{$({\textit{yld}}_1, {\textit{yld}}_2, {\textit{yld}}_3)$} & \multicolumn{1}{c|}{$-10$} & \multicolumn{1}{c|}{$-10$} & \multicolumn{1}{c|}{$-10$} \\ \hline
\end{tabular}}
\vspace*{0.2cm}
\end{subfigure}
\vspace*{-0.2cm}
\caption{Example: Cars at an intersection and the corresponding NFG.}\label{intersection:fig}
\vspace*{-0.6cm}
\end{figure}
\begin{examp} Consider the scenario, based on an example from~\cite{EP14}, where three cars meet at an intersection and want to proceed as indicated by the arrows in \figref{intersection:fig}. Each car can either \emph{proceed} or \emph{yield}. If two cars with intersecting paths proceed, then there is an accident. If an accident occurs, the car having the right of way, i.e., the other car is to its right, has a utility of $-100$ and the car that should yield has a utility of $-1000$. If a car proceeds without causing an accident, then its utility is $5$ and the cars that  yield have a utility of $-5$. If all cars yield, then, since this delays all cars, all have utility $-10$. The 3-player NFG is given in~\figref{intersection:fig}. Considering the different optimal equilibria of the NFG:
\begin{itemize}
\item
the SWNE and SWCE are the same: for $c_2$ to yield and $c_1$ and $c_3$ to proceed, with the expected utilities $(5,-5,5)$;
\item
the SFNE is for $c_1$ to yield with probability $1$, $c_2$ to yield with probability $0.863636$ and $c_3$ to yield with probability $0.985148$, with the expected utilities $(-9.254050, -9.925742, -9.318182)$; 
\item
the SFCE gives a joint distribution where the probability of $c_2$ yielding and of $c_1$ and $c_3$ yielding are both $0.5$ with the expected utilities $(0,0,0)$. 
\end{itemize}
Modifying $u_2$ such that $u_2(\textit{pro}_1,\textit{pro}_2,\textit{pro}_3)=-4.5$ to, e.g., represent a reckless driver, the SWNE becomes for $c_1$ and $c_3$ to yield and $c_2$ to proceed with the expected utilities $(-5,5,-5)$, while
the SWCE is still for $c_2$ to yield and $c_1$ and $c_3$ to proceed. 
The SFNE and SFCE also do not change.
\end{examp}

\startpara{Algorithms for computing equilibria} 
Before we give our algorithm to compute correlated equilibria, we briefly describe the approach of \cite{KNPS20b,KNPS21} for Nash equilibria computation that this paper builds upon. 
Finding NE in two-player NFGs is in the class of \emph{linear complementarity} problems (LCPs) and we follow the algorithm presented in \cite{KNPS21}, which reduces the problem to SMT via labelled polytopes~\cite{NRTV07} by considering the regions of the strategy profile space, iteratively reducing the search space as positive probability assignments are found and added as restrictions on this space. To find SWNE and SFNE, we can enumerate all NE and then find the optimal NE.

When there are more than two players, computing NE values becomes a more complex task, as finding NE within a given support no longer reduces to a linear programming (LP) problem. In~\cite{KNPS20b} we presented an algorithm using support enumeration~\cite{PNS04}, which exhaustively examines all sub-regions, i.e., supports, of the strategy profile space, one at a time, checking whether that sub-region contains NEs. For each support, finding SWNE can be reduced to a \emph{nonlinear programming problem}~\cite{KNPS20b}. This nonlinear programming problem can be modified to find SFNE in each support, similarly to how the LP problem for SWCEs is modified to find SFCEs below.

In the case of CE we can first find a joint strategy for the players, i.e., a distribution over the action tuples, which, as explained above, can then be mapped to a correlated profile. A SWCE can be found by solving the following LP problem.  Maximise: $\sum_{i \in N}\sum_{\alpha \in A} u_i(\alpha) \cdot p_\alpha$ subject to:
\begin{eqnarray}
\mbox{$\sum_{\alpha_{-i} \in A_{-i}}(u_i(\alpha_{-i}[a_i])-u_{i}(\alpha_{-i}[a_i'])) \cdot p_{\alpha_{-i}[a_i]} \geq 0$} \label{eq:corr1} \\
\mbox{$0 \leq p_\alpha \leq 1$} \label{eq:corr2} \\
\mbox{$\sum_{\alpha \in A} p_\alpha = 1$} \label{eq:corr3}
\end{eqnarray}
for all $i \in N$, $\alpha \in A$, $a_i, a_i' \in A_i$, $\alpha_{-i} \in A_{-i}$ where $A_{-i} \rmdef \{ \alpha_{-i} \mid \alpha \in A \}$. The variables $p_\alpha$ represent the probability of the joint strategy corresponding to the correlated profile selecting the action-tuple $\alpha$.  
The above LP has $|A|$ variables, one for each action-tuple, and $\sum_{i \in N}(|A_i|^2-|A_i|) + |A| + 1$ constraints.
Computation of SFCE can be reduced to the following optimisation problem. Minimise $p^{\max} - p^{\min}$ subject to: \eqnref{eq:corr1}, \eqnref{eq:corr2} and \eqnref{eq:corr3} together with:
\begin{eqnarray}
\mbox{$p^i = \mbox{$\sum_{\alpha \in A}$} p_\alpha\cdot u_i(\alpha)$} \label{eq:frpayoff} \\
\mbox{$\left( \wedge_{m \in N} p^i \geq p^m \right) \rightarrow (p^{\max} = p^i$}) \label{eq:frpmax} \\
\mbox{$\left( \wedge_{m \in N} p^i \leq p^m \right) \rightarrow (p^{\min} = p^i$}) \label{eq:frpmin} 
\end{eqnarray}
for all $i \in N$, $m \neq i$, $\alpha \in A$, $a_j, a_l \in A_i$, $\alpha_{-i} \in A_{-i}$. Again, the variables $p_\alpha$ in the program represent the probability of the players playing the joint action $\alpha$.  The constraint \eqnref{eq:frpayoff} requires $p^i$ to equal the utility of player $i$. The constraints \eqnref{eq:frpmax} and \eqnref{eq:frpmin} set $p^{\max}$ and $p^{\min}$ as the maximum and minimum values within the utilities of the players, respectively. Given we use the constraints \eqnref{eq:corr1}, \eqnref{eq:corr2} and \eqnref{eq:corr3}, we start with the same number of variables and constraints as needed to compute SWCEs and incur an additional $|N|+2$ variables and $3\cdot|N|$ constraints. 






\startpara{Implementation} To find SWNE or SFNE of two-player NFGs,
we adopt a similar approach to \cite{KNPS21},
using labelled polytopes to characterise and find NE values through a reduction to SMT in both Z3~\cite{Z3} and Yices~\cite{Dut14}.
As an optimised precomputation step, when possible we also search for and filter out \emph{dominated strategies}, which speeds up the computation and reduces solver calls.

For NFGs with more than two players, solving the nonlinear programming problem based on support enumeration has been implemented in \cite{KNPS20b} using a combination of the SMT solver Z3~\cite{Z3} and the nonlinear optimisation suite {\sc Ipopt}~\cite{Wac09}. To mitigate the inefficiencies of an SMT solver for such problems, we used Z3 to filter out unsatisfiable support assignments with a timeout and then {\sc Ipopt} is called to find SWNE values using an interior-point filter line-search algorithm~\cite{WB06}. To speed up the overall computation, the support assignments are analysed in parallel. Computing SFNE increases the complexity of the nonlinear program and, due to the inefficiency in this approach~\cite{KNPS20b}, we have not extended the implementation to compute SFNE.

As shown above, computing SWCE for NFGs reduces to solving an LP, and we implement this using either the optimisation solver Gurobi \cite{gurobi} or the SMT solver Z3 \cite{Z3}. In the case of SFCE, the constraints \eqnref{eq:frpmax} and \eqnref{eq:frpmin} include implications, and therefore the problem does not reduce directly to an LP. When using Z3, we can encode these constraints directly as it supports assertions that combine inequalities with logical implications, a feature that linear solvers such as Gurobi do not have. \sectref{conc-sect} discusses implementing SFCE computation in Gurobi. Both solvers support the specification of \emph{lower priority} or \emph{soft} objectives, which makes it possible to have a consistent ordering for the players' payoffs in cases where multiple equilibria exist.



\begin{table}[t]
\centering
{\scriptsize
\begin{tabular}{|c|r|r|r|r|r|r|r|r|} \hline
\multirow{2}{*}{Game} & \multicolumn{1}{c|}{\multirow{2}{*}{Players}} & \multicolumn{1}{c|}{\multirow{2}{*}{$|A_i|$}} & \multicolumn{1}{c|}{\multirow{2}{*}{$|A|$}} & 
\multicolumn{2}{c|}{\multirow{1}{*}{NE}} & 
\multicolumn{2}{c|}{CE} \\ \cline{5-8}
& & & & \multicolumn{1}{c|}{\multirow{1}{*}{Supports}} & \multicolumn{1}{c|}{SW} & \multicolumn{1}{c|}{SW} & \multicolumn{1}{c|}{SF} 
\\ 
\hline
\multirow{6}{*}{\shortstack[c]{\emph{Majority voting} \\ \emph{games}}}
&  \multirow{5}{*}{2}  &  4 &    16 &       225 &  0.07 & 0.02 & 0.08 \\
&                      &  6 &    36 &     3,969 &   0.1 & 0.02 & 0.1 \\
&                      &  8 &    64 &    65,025 &   0.4 & 0.03 & 0.3 \\
&                      & 10 &   100 & 1,046,529 &   5.8 & 0.07 & 0.7 \\ \cline{2-8}
&  \multirow{2}{*}{3}  &  3 &    27 &       343 &  1.2 & 0.07 & 0.1 \\
&                      &  4 &    81 &     3,375 & 25.8 & 0.08 & 0.3 \\ \cline{2-8}
\hline\hline
\multirow{5}{*}{\shortstack[c]{\emph{Covariant} \\ \emph{games}}}
& \multirow{2}{*}{3}  & 3 &    27 &       343 &    8.7 & 0.08 & 1.7 \\
&                     & 4 &    81 &     3,375 &  598.5 & 0.08 & 2.9 \\ \cline{2-8}
& \multirow{2}{*}{8}  & 2 &   256 &     6,561 &  TO &  0.3 & TO \\
&                     & 3 & 6,561 & 5,764,801 &  TO & 22.8 & TO \\ \cline{2-8}
& \multirow{1}{*}{10} & 2 & 1,024 &    59,049 &  TO &  1.2 & TO \\
\hline
\end{tabular}}
\vspace*{0.1cm}
\caption{Times for synthesis of equilibria in NFGs (timeout 30 mins).}\label{tab:normal_form}
\vspace*{-0.8cm}
\end{table}

\startpara{Efficiency and scalability}
\tabref{tab:normal_form} presents experimental results for solving a selection of NFGs
randomly generated with GAMUT~\cite{NWSL04}, using Gurobi for SWCE and NE of two-player NFGs, Z3 for SFCE and both {\sc Ipopt} and Z3 for NFGs of more than two players,
and running on a 2.10GHz Intel Xeon Gold with 32GB of JVM memory.
For each instance, \tabref{tab:normal_form} lists the number of players, actions for each player, joint actions and supports that need to be enumerated when finding NE, as well as the time to find SWNEs, SWCEs and SFCEs (the time for finding SFNEs of two-player games is the same as for SWNEs). As the results demonstrate, due to a simpler problem being solved and the fact that we do not need to enumerate the solutions, computing CEs scales far better than NEs as the number of players and actions increases. Finding NEs in games with more than two players is particularly hard as the constraints are nonlinear.
We also see that SFCE computation is slower than SWCE, which is caused by the additional variables and constraints required when finding SFCE and using Z3 rather than Gurobi for the solver.


\section{Concurrent Stochastic Games}\label{csg-sect}

We now further develop our approach to support
concurrent stochastic games (CSGs)~\cite{Sha53}, in which players repeatedly make simultaneous action choices that cause the game's state to be updated probabilistically. We extend the previously introduced definitions of optimal equilibria to such games, focusing on 
subgame-perfect equilibria,  which are equilibria in every state of a CSG. We then present 
algorithms to reason about and synthesise such equilibria. 

\begin{definition}[Concurrent stochastic game]\label{def:csg}
A \emph{concurrent stochastic multi-player game} (CSG) is a tuple
$\game = (N, S, \bar{S}, A, \Delta, \delta, \AP, \lab)$ where:
\begin{itemize}
\item $N=\{1,\dots,n\}$ is a finite set of players;
\item $S$ is a finite set of states and $\bar{S} \subseteq S$ is a set of initial states;
\item $A = (A_1\cup\{\bot\}) {\times} \cdots {\times} (A_n\cup\{\bot\})$ and $A_i$ is a finite set of actions available to player $i \in N$ and $\bot$ is an idle action disjoint from the set $\cup_{i=1}^n A_i$;
\item $\Delta \colon S \rightarrow 2^{\cup_{i=1}^n A_i}$ is an action assignment function;
\item $\delta \colon (S {\times} A) \rightarrow \dist(S)$ is a (partial) probabilistic transition function;
\item $\AP$ is a set of atomic propositions and $\lab \colon S \rightarrow 2^{\AP}$ is a labelling function.
\end{itemize}
\end{definition}
For the remainder of this section we fix a CSG $\game$ as in \defref{def:csg}.
The game $\game$ starts in one of its initial states $\sinit \in \bar{S}$ and, supposing $\game$ is in a state $s$, then each player $i$ of $\game$ chooses an action from the set that are available, defined as $A_i(s) \rmdef \Delta(s) \cap A_i$ if $\Delta(s) \cap A_i$ is non-empty and $A_i(s) \rmdef \{\bot\}$ otherwise. Supposing each player chooses $a_i$, then the game transitions to state $s'$ with probability $\delta(s,(a_1,\dots,a_n))$. To enable quantitative analysis of $\game$ we augment it with \emph{reward structures}, which are tuples $r{=}(r_A,r_S)$ of an action reward function $r_A \colon S {\times} A \ra \Rset$ and state reward function $r_S \colon S \ra \Rset$.

A \emph{path} of $\game$ is a sequence $\pi = s_0 \xrightarrow{\alpha_0} s_1 \xrightarrow{\alpha_1} \cdots$ where $s_k \in S$, $\alpha_k = (a^k_1,\dots,a^k_n) \in A$, $a^k_i \in A_i(s_k)$ for $i \in N$ and $\delta(s_k,\alpha_k)(s_{k+1})>0$ for all $k \geq 0$. 
We denote by $\fpaths_{\game,s}$ and $\ipaths_{\game,s}$ the sets of finite and infinite paths starting in state $s$ of $\game$ respectively and drop the subscript $s$ when considering all finite and infinite paths of $\game$. As for NFGs, we can define \emph{strategies} of $\game$ that resolve the choices of the players. Here, a strategy for player $i$ is a function $\sigma_i \colon \fpaths_{\game} \ra \dist(A_i \cup \{ \bot \})$ such that, if $\sigma_i(\pi)(a_i){>}0$, then $a_i \in A_i(\last(\pi))$ where $\last(\pi)$ is the final state of $\pi$. Furthermore, we can define strategy profiles, correlated profiles and joint strategies analogously to \defdefref{strats-nfgs}{corp-nfgs}. 

The utility of a player $i$ of $\game$ is defined by a random variable $X_i \colon \ipaths_{\game} \rightarrow \Rset$ over infinite paths. For a profile\footnote{We can also construct such a probability measure and expected value given a correlated profile or joint strategy.} $\sigma$ and state $s$, using standard techniques~\cite{KSK76}, we can construct a probability measure $\Prob^{\sigma}_{\game,s}$ over the paths with initial state $s$ corresponding to $\sigma$, denoted $\ipaths^\sigma_{\game,s}$ and the expected value $\Eset^{\sigma}_{\game,s}(X_i)$ of player $i$'s utility from $s$ under $\sigma$. Given utilities $X_1,\dots,X_n$ for all the players of $\game$, we can then define NE and CE (see \defref{def:eq}) as well as the restricted classes of SW and SF equilibria as for NFGs (see \defref{def:swne}).
Following \cite{KNPS21,KNPS20b}, we focus on \emph{subgame-perfect} equilibria~\cite{OR04}, which are equilibria in \emph{every state} of $\game$.

\startpara{Nonzero-sum properties} As in \cite{KNPS21} (for two-player CSGs) and \cite{KNPS20b} (for $n$-player CSGs) we can specify equilibria-based properties using temporal logic. For simplicity, we restrict attention to nonzero-sum properties without nesting, allowing for the specification of NE and CE against either SW or SF optimality.

\vspace*{-0.05em}
\begin{definition}[Nonzero-sum specifications]\label{def:rpatlsyn}
The syntax of nonzero-sum specifications $\theta$ for CSGs is given by the grammar:
\begin{eqnarray*}
\phi & \; \coloneqq \; &  \nashop{\mathbb{C}}{\star_1,\star_2}{\opt \sim x}{\theta} \\
\theta & \; \coloneqq \; & \probop{}{\psi}{+}{\cdots}{+}\probop{}{\psi} \ \mid \ \rewop{r}{}{\rho}{+}{\cdots}{+}\rewop{r}{}{\rho} \\
\psi & \; \coloneqq \; & \next \ap \ \mid \ \ap \buntil \ap \ \mid \ \ap \until \ap \\
\rho & \; \coloneqq \; &  \sinstant{=k} \ \mid \ \scumul{\leq k} \ \mid \ \future \ap
\end{eqnarray*}
where $\mathbb{C} = C_1{:}\cdots{:}C_m$, $C_1,\dots,C_m$ are coalitions of players such that $C_i \cap C_j = \emptyset$ for all $1\leq i \neq j \leq m$ and $\cup_{i=1}^m C_i = N$,  $(\star_1,\star_2) \in \{\NE,\CE \} {\times} \{\SW, \SF \}$, $\opt \in \{ \min,\max\}$, $\sim \,\in \{<, \leq, \geq, >\}$, $x \in \Qset$, $r$ is a reward structure,  $k \in \Nset$ and $\ap$ is an atomic proposition. %
\end{definition}
The nonzero-sum formulae of \defref{def:rpatlsyn} extend the logic of in \cite{KNPS21,KNPS20b} in that we can now specify the type of equilibria, NE or CE, and optimality criteria, SW or SF.
A probabilistic formula $\nashop{C_1{:}{\cdots}{:}C_m}{\star_1,\star_2}{\max \sim x}{\probop{}{\psi_1}{+}{\cdots}{+}\probop{}{\psi_m}}$ is true in a state if, when the players form the coalitions $C_1,\dots,C_m$,
there is a subgame-perfect equilibrium of type $\star_1$ meeting the optimality criterion $\star_2$  for which the \emph{sum} of the values of the objectives
$\probop{}{\psi_1},\dots,\probop{}{\psi_m}$ for the coalitions $C_1,\dots,C_m$
satisfies ${\sim} x$. The objective $\psi_i$ of coalition $C_i$ is either a next ($\next \ap$), bounded until ($\ap_1 \buntil  \ap_2$) or until ($\ap_1 \until \ap_2$) formula, with the usual equivalences, e.g., $\future\ap \equiv \true \until \ap$.

For a reward formula $\nashop{C_1{:}{\cdots}{:}C_m}{\star_1,\star_2}{\opt \sim x}{\rewop{r_1}{}{\rho_1}{+}{\cdots}{+}\rewop{r_m}{}{\rho_m}}$ the meaning is similar; however, here the objective of coalition $C_i$ refers to a reward formula $\rho_i$ with respect to reward structure $r_i$ and this formula is either a bounded instantaneous reward ($\sinstant{=k}$), bounded accumulated reward ($\scumul{\leq k}$) or reachability reward ($\future \ap$).

For formulae of the form $\nashop{C_1{:}{\cdots}{:}C_m}{\star_1,\star_2}{\min \sim x}{\theta}$, the dual notions of cost equilibria are considered.
We also allow \emph{numerical} queries of the form $\nashop{C_1{:}{\cdots}{:}C_m}{\star_1,\star_2}{\opt =?}{\theta}$, which return the sum of the optimal subgame-perfect equilibrium's values.
\startpara{Model checking nonzero-sum specifications} Similarly to \cite{KNPS21,KNPS20b}, to allow model checking of nonzero-sum properties we consider a restricted class of CSGs.
We make the following assumption, which can be checked using graph algorithms with time complexity quadratic in the size of the state space~\cite{dA97a}.
\begin{assumption}\label{game-assum}
For each subformula $\probop{}{\ap_1 \until \ap_2}$, a state labelled $\neg \ap_1 \vee \ap_2$ is reached with probability 1 from all states under all strategy profiles and correlated profiles. For each subformula $\rewop{r}{}{\future \ap}$, a state labelled $\ap$ is reached with probability 1 from all states under all strategy profiles and correlated profiles.
\end{assumption} 
We now show how to compute the optimal values of a nonzero-sum formula $\phi = \nashop{C_1{:}{\cdots}:C_m}{\star_1,\star_2}{\opt\sim x}{\theta}$ when $\opt=\max$. The case when $\opt = \min$ can be computed by
negating all utilities and maximising.

The model checking algorithm broadly follows those presented in \cite{KNPS21,KNPS20b}, with the differences described below. The problem is reduced to solving an $m$-player \emph{coalition game} $\game^\cC$ where $\cC = \{C_1,\dots,C_m\}$ and the choices of each player $i$ in $\game^\cC$ correspond to the choices of the players in coalition $C_i$ in $\game$. Formally, we have the 
following definition in which, without loss of generality, we assume $\cC$ is of the form $\{ \{1,\dots,n_1\}, \{ n_1{+}1,\dots n_2 \}, \dots, \{ n_{m-1}{+}1, \dots n_{m} \}\}$ and let $j_\cC$ denote player $j$'s position in its coalition. 
\begin{definition}[Coalition game] For CSG $\game =  (N, S, \bar{S}, A, \Delta, \delta, \AP, \lab)$ and partition $\cC=\{C_1, \dots, C_{m}\}$ of the players into $m$ coalitions, we define the \emph{coalition game} $\game^\cC = ( \{1,\dots,m\}, S, \bar{S}, A^\cC, \Delta^\cC, \delta^\cC, \AP, \lab)$ as an $m$-player CSG where:
\begin{itemize}
	\item $A^\cC = (A^\cC_1\cup \{ \bot\}) {\times} \cdots {\times} (A^\cC_{m}\cup \{ \bot\})$;
	\item $A^\cC_i = (\prod_{j \in C_i} (A_j\cup\{\bot\}) \setminus \{(\bot,\dots,\bot)\} \big)$ for all $1\leq i\leq m$;
	\item for any $s \in S$ and $1\leq i\leq m\!:$ $a_i^\cC \in \Delta^\cC(s)$ if and only if either $\Delta(s) \cap A_j =\emptyset$ and $a_i^\cC(j_\cC)=\bot$ or $a_i^\cC(j_\cC) \in \Delta(s)$  for all $j \in C_i$;
	\item for any $s \in S$ and $(a^\cC_1,\dots,a^\cC_{m})\in A^\cC\!:$ $\delta^\cC(s,(a^\cC_1,\dots,a^\cC_{m})) = \delta(s,(a_1,\dots,a_n))$ where for $i \in M$ and $j \in C_i$ if $a_i^\cC {=} \bot$, then $a_j {=}\bot$ and otherwise $a_j {=} a_i^\cC(j_\cC)$.
\end{itemize}
\end{definition}
If all the objectives in $\theta$ are finite-horizon,  \emph{backward induction}~\cite{SW+01,NMK+44} can be applied to compute (precise) optimal equilibria values with respect to the criterion $\star_2$ and equilibria type $\star_1$. On the other hand, if all the objectives are infinite-horizon, \emph{value iteration}~\cite{CH08} can be used to approximate optimal equilibria values and, when there is a combination of objectives, the game under study is modified in a standard manner to make all objectives infinite-horizon. 

Backward induction and value iteration over the CSG $\game^\cC$
both work by iteratively computing new values for each state $s$ of $\game^\cC$.
The values for each state, in each iteration,
are found by computing optimal equilibria values of an NFG $\nfgame$
whose utility function is derived from the outgoing transition probabilities from $s$ in the CSG
and the values computed for successor states of $s$ in the previous iteration.
%
%
The difference here, with respect to \cite{KNPS20b},
is that the NFGs are solved for the additional equilibria and optimality conditions
considered in this paper,
which we compute using the algorithms presented in \sectref{nfg-sect}.

\startpara{Algorithm for probabilistic until}
We present the details
of value iteration for (unbounded) probabilistic until,~i.e.,
for $\phi = \nashop{C_1{:}{\cdots}:C_m}{\star_1,\star_2}{\max\sim x}{\theta}$
where $\theta = \probop{}{\ap^1_1 \until \ap^1_2}{+}\cdots{+}\probop{}{\ap^{m}_1 \until \ap^{m}_2}$ and the complete model checking algorithm can be found in \appref{appendix}.

Following \cite{KNPS20b}, we use $\V_{\game^\cC}(s,\star_1,\star_2,\theta,n)$
to denote the vector of computed values, at iteration $n$,
in state $s$ of $\game^\cC$ for optimality criterion $\star_2$ (SW or SF),
equilibria type $\star_1$ (NE or CE) and (until) objectives $\theta$.
We also use $\mathbf{1}_m$ and $\mathbf{0}_m$ to denote a vector of size $m$ whose entries all equal to 1 or 0, respectively.
For any set of states $S'$, atomic proposition $\ap$ and state $s$ we let $\eta_{S'}(s)$ equal $1$ if $s \in S'$ and $0$ otherwise, and $\eta_{\ap}(s)$ equal $1$ if $\ap \in L(s)$ and $0$ otherwise.

Each step of value iteration also keeps track of two sets $D,E\subseteq M$,
where $M = \{1,\dots,m\}$ are the players of $\game^\cC$.
We use $D$ for the subset of players that have already reached their goal (by satisfying $\ap_2^i$)
and $E$ for the players who can no longer can satisfy their goal
(having reached a state that fails to satisfy $\ap_1^i$).
It can then be ensured that their payoffs no longer change and are set to 1 or 0, respectively.
In these cases, we effectively consider a modified game where,
although the payoffs for these players are set,
we still need to take their strategies into account in order to guarantee an optimal equilibrium.



Optimal values for all states $s$ in the CSG $\game^\cC$ can be computed as the following limit: $\V_{\game^\cC}(s,\star_1,\star_2,\theta) = \lim_{n \ra \infty} \V_{\game^\cC}(s,\star_1,\star_2,\theta,n)$, where $\V_{\game^\cC}(s,\star_1,\star_2,\theta,n)=\V_{\game^\cC}(s,\star_1,\star_2,\emptyset,\emptyset,\theta,n)$ and, for any $D,E \subseteq M$ such that $D \cap E = \emptyset$:
\[
\V_{\game^\cC}(s,\star_1,\star_2,D,E,\theta,n) = \left\{ \begin{array}{cl}
(\eta_{D}(1),\dots,\eta_{D}(m)) & \;\; \mbox{if $D \cup E = M$} \\
(\eta_{\ap^1_2}(s),\dots,\eta_{\ap^{m}_2}(s)) & \;\; \mbox{else if $n=0$} \\
\V_{\game^\cC}(s,\star_1,\star_2,D \cup D',E,\theta,n) & \;\; \mbox{else if $D' \neq \varnothing$} \\
\V_{\game^\cC}(s,\star_1,\star_2,D,E \cup E',\theta,n) & \;\; \mbox{else if $E' \neq \varnothing$} \\
\val(\nfgame,\star_1,\star_2) & \;\; \mbox{otherwise}
\end{array} \right. 
\]
where $D' =  \{l \in M {\setminus} (D\cup E) \mid \ap^l_2 \in L(s) \}$, $E' =  \{l \in M {\setminus} (D\cup E) \mid \ap^l_1 \not\in L(s) \; \mbox{and} \; s \in L( \ap^l_2) \}$ and $\val(\nfgame,\star_1,\star_2)$ equals optimal values of the NFG $\nfgame = (M,A^\cC,u)$  with respect to the criterion $\star_2$ and of equilibria type $\star_1$ in which for any $1{\leq}l{\leq}m$ and $\alpha \in A^\cC$:
\[ 
u_l(\alpha) = \left\{
\begin{array}{cl}
1 & \;\;\mbox{if $l \in D$} \\
0 & \;\;\mbox{else if $l \in E$} \\
\sum_{s' \in S} \delta^{\cC}(s,\alpha)(s') \cdot v^{s',l}_{n-1} & \;\;\mbox{otherwise}
\end{array} \right.
\]
and $(v^{s',1}_{n-1},v^{s',2}_{n-1},\dots,v^{s',m}_{n-1}) = \V_{\game^\cC}(s',\star_1,\star_2,D,E,\theta,n{-}1)$ for all $s' \in S$.

\vskip6pt
Since this paper considers equilibria for any number of coalitions
(in particular, for more than two),
the above follows the algorithm of~\cite{KNPS20b}
in the way that it keeps track of the coalitions that
have satisfied their objective ($D$) or can no longer do so ($E$).
By contrast the CSG algorithm of \cite{KNPS21} was limited to two coalitions,
which enabled the exploitation of efficient MDP analysis techniques
for such coalitions.
%
As explained in \cite{KNPS20b}, in such a scenario we cannot reduce the analysis from an $n$-coalition game to an $(n-1)$-coalition game, as otherwise we would 
give one of the remaining coalitions additional power (the action choices of the coalition that has satisfied their objective or can no longer do so), which would therefore give this coalition an advantage over the other coalitions.

\startpara{Strategy synthesis} As in \cite{KNPS21,KNPS20b} we can extend the model checking algorithm to perform \emph{strategy synthesis}, generating a witness (i.e., a profile or joint strategy) representing the corresponding optimal equilibrium. This is achieved by storing the profile or joint strategy for the NFG solved in each state. Both the profiles and joint strategies require finite memory and are probabilistic. Memory is required as choices change after a path formula becomes true or a target is reached and to keep track of the step bound in finite-horizon properties. Randomisation is required for both NE and CE of NFGs.

\startpara{Correctness and complexity} The correctness of the algorithm follows directly from~\cite{KNPS21,KNPS20b}, as changing the class of equilibria or optimality criterion does not change the proof. The complexity of the algorithm is linear in the formula size and value iteration requires finding optimal NE or CE for an NFG in each state of the model. Computing NEs of an NFG with two (or more) players is PPAD-complete \cite{DGP09,CDT09}, while finding optimal CEs of an NFG is in P~\cite{GZ89}.

\section{Case Studies and Experimental Results}\label{sec:results}

We have developed an implementation of our techniques for equilibria synthesis on CSGs,
described above, building on top of the PRISM-games~\cite{KNPS20} model checker.
Our implementation extends the tool's existing support for
construction and analysis of CSGs, which is contained within
its sparse matrix based ``explicit'' engine written in Java.
We have considered a range of CSG case studies
(supplementary material can be found at~\cite{www}).
Below, we summarise the efficiency and scalability of our approach,
again running on a 2.10GHz Intel Xeon Gold with 32GB JVM memory,
and then describe our findings on individual case studies.

\begin{table}[!t]
\centering
{\scriptsize
\begin{tabular}{|c|c|c|c|r|r|r|r|r|} \hline
\multicolumn{1}{|c|}{Case study \& property} & 
\multicolumn{1}{c|}{\multirow{2}{*}{Players}} &
\multicolumn{1}{c|}{\multirow{2}{*}{$\star_1{,}\star_2$}} &
\multicolumn{1}{c|}{Param.} & 
\multicolumn{2}{c|}{CSG statistics} & \multicolumn{1}{c|}{Constr.} & \multicolumn{1}{c|}{Verif.} \\ 
\cline{5-8}
\multicolumn{1}{|c|}{[parameters]} &
& &
\multicolumn{1}{c|}{values} & 
\multicolumn{1}{c|}{States} & 
\multicolumn{1}{c|}{Trans.} & \multicolumn{1}{c|}{time(s)} &  \multicolumn{1}{c|}{time (s)} \\ \hline \hline
\multirow{7}{*}{\shortstack[c]{\emph{Aloha} \\
$({\star_1{,}\star_2})_{\min{{=}?}}(\rewop{\mathit{time}}{}{\future \mathsf{s}_i})$ \\
$\mbox{[$b_{\mathit{max}},\mathit{q}$]}$}}
& \multirow{4}{*}{2} & NE,SW & \multirow{4}{*}{4,0.8} & \multirow{4}{*}{2,778} & \multirow{4}{*}{6,285} & \multirow{4}{*}{0.1} & 2.2  \\
& & CE,SW & & & & &  2.1 \\
& & NE,SF & & & & &  2.1 \\
& & CE,SF & & & & & 23.3 \\ \cline{2-8}
& \multirow{2}{*}{3} & CE,SW & \multirow{2}{*}{4,0.8} & \multirow{2}{*}{107,799} & \multirow{2}{*}{355,734} & \multirow{2}{*}{3.0} & 80.1 \\
& & CE,SF & & & & & 114.6 \\ \cline{2-8}
& \multirow{2}{*}{4} & NE,SW & \multirow{2}{*}{2,0.8} & \multirow{2}{*}{68,689} & \multirow{2}{*}{161,904} & \multirow{2}{*}{1.9} & 1042.9 \\
& & CE,SW & & & & & 58.8
\\
\hline\hline
\multirow{4}{*}{\shortstack[c]{\emph{Aloha} \\
$({\star_1{,}\star_2})_{\max{{=}?}}(\probop{\max{=}?}{\future \mathsf{s}_i {\wedge} t{\leq}D})$ \\
$\mbox{[$b_{\mathit{max}},\mathit{q}, \mathit{D}$]}$}}
& \multirow{2}{*}{4} & NE,SW & \multirow{2}{*}{2,0.8,8} & \multirow{2}{*}{159,892} & \multirow{2}{*}{388,133} & \multirow{2}{*}{3.9} & 1027.5  \\ 
&   & CE,SW & & & & & 224.5  \\ \cline{2-8}
& \multirow{2}{*}{5} & CE,SW & \multirow{2}{*}{2,0.8,8} & \multirow{2}{*}{1,797,742} & \multirow{2}{*}{5,236,655} & \multirow{2}{*}{54.5} & 4,936.8  \\
& & CE,SF & & &    & & TO \\
\hline\hline
\multirow{4}{*}{\shortstack[c]{\emph{Power control} \\
$({\star_1{,}\star_2})_{\max{{=}?}}(\rewop{\mathit{r}}{}{\future \mathsf{e}_i})$ \\
$\mbox{[$pow_{\mathit{max}},e_{\mathit{max}}, q_{\mathit{fail}}$]}$}} 
& \multirow{3}{*}{2} & NE,SW & \multirow{3}{*}{8,40,0.2} &  \multirow{3}{*}{32,812} &  \multirow{3}{*}{260,924} & \multirow{3}{*}{1.2} & 564.5 \\
& & NE,SF & & &    & & 566.3  \\ 
& & CE,SW & & &    & & 177.9  \\ \cline{2-8}
& \multirow{2}{*}{3} & CE,SW & \multirow{2}{*}{5,15,0.2} & \multirow{2}{*}{42,156} & \multirow{2}{*}{740,758} & \multirow{2}{*}{3.5} & 147.0  \\
& & CE,SF & & &    & & TO  \\
\hline\hline
\multirow{4}{*}{\shortstack[c]{\emph{Public good} \\
$({\star_1{,}\star_2})_{\max{{=}?}}(\rewop{\mathit{c}}{}{\sinstant{=r_\mathit{max}}})$ \\
$\mbox{[$f,r_{\mathit{max}}$]}$}} 
& \multirow{2}{*}{3} & NE,SW & \multirow{2}{*}{2.5,3} & \multirow{2}{*}{16,202} & \multirow{2}{*}{35,884} & \multirow{2}{*}{0.8} & 27.5 \\
& & CE,SW & & &    & & 1.9  \\ \cline{2-8}
& \multirow{2}{*}{4} & NE,SW & \multirow{2}{*}{3,3} & \multirow{2}{*}{391,961} & \multirow{2}{*}{923,401} & \multirow{2}{*}{13.0} & 71.9 \\
& & CE,SW & & &    & & 35.3 \\ \cline{2-8}
& 5 & CE,SW & 4,2 & 59,294 & 118,342 & 3.1 & 5.2  \\
\hline\hline
\multirow{4}{*}{\shortstack[c]{\emph{Investors} \\
$({\star_1{,}\star_2})_{\max{{=}?}}(\rewop{\mathit{prof}}{}{\future \mathsf{cin}_i})$ \\
$\mbox{[$p_{\mathit{bar}},\mathit{months}$]}$}} 
& \multirow{2}{*}{2} & CE,SW & \multirow{2}{*}{0.2,8} & \multirow{2}{*}{71,731} & \multirow{2}{*}{315,804} & \multirow{2}{*}{2.4} & 47.5 \\
& & CE,SF & & &    & & 2,401.9  \\ \cline{2-8}
& \multirow{2}{*}{3} & CE,SW & \multirow{2}{*}{0.2,5} & \multirow{2}{*}{83,081} & \multirow{2}{*}{462,920} & \multirow{2}{*}{3.6} & 79.3  \\
& & CE,SF & & &    & & 861.2  \\
\hline
\end{tabular}}
\vspace*{0.1cm}
\caption{Statistics for a set of CSG verification instances (timeout 2 hours).}
\label{tab:case_studies}
\vspace{-1cm}
\end{table}
\startpara{Efficiency and scalability}
\tabref{tab:case_studies} summarises the performance of our implementation
on the case studies that we have considered. It shows the statistics for each CSG,
and the time taken to build it and perform equilibria synthesis,
for several different variants (NE vs. CE, SW vs. SF).
Comparing the efficiency of synthesising SWNE and SWCE,
we see that the latter is typically much faster.
For two-player NE, the social fairness variant is no more expensive to compute as we enumerate all NEs. For CE, which uses Z3 rather than Gurobi for finding SF, we note that, although Z3 is able to find optimal equilibria, it is not primarily developed as an optimisation suite, and therefore generally performs poorly in comparison with Gurobi.
The benefits of the social fair equilibria, in terms of the values yielded for individual players,
are discussed in the in-depth coverage of the different case studies below.

\startpara{Aloha} In this case study, introduced in~\cite{KNPS21}, a number of users try to send packets using the slotted Aloha protocol. We suppose that each user has one packet to send and, in a time slot, if $k$ users try and send their packet, then the probability that each packet is successfully sent is $q/k$ where $q \in [0,1]$.
If a user fails to send a packet, then the number of slots it waits before resending the packet is set according to Aloha's exponential backoff scheme. The scheme requires that each user maintains a backoff counter, which it increases each time there is a packet failure (up to $b_{\max}$) and, if the counter equals $k$ and a failure occurs, randomly chooses the slots to wait from $\{0,1,\dots,2^k{-}1\}$. 

We suppose that the objective of each user is to minimise the expected time to send their packet, which is represented by the nonzero-sum formula $\nashop{\mathit{usr}_1{:}\cdots{:}\mathit{usr}_m}{\star_1,\star_2}{\min=?}{\rewop{\mathit{time}}{}{\future \mathsf{s}_1}{+}{\cdots}{+}\rewop{\mathit{time}}{}{\future \mathsf{s}_m}}$. Synthesising optimal strategies for this specification, we find that the cases for SWNE and SWCE coincide (although 
SWCE 
returns a joint strategy for the players, this joint strategy can be separated to form a strategy profile). This profile requires one user to try and send first,  and then for the remaining users to take turns to try and send afterwards. If a user fails to send, then they enter backoff and allow all remaining users to try and send before trying to send again. There is no gain to a user in trying to send at the same time as another, as this will increase the probability of a sending failure, and therefore the user having to spend time in backoff before getting to try again. For SFNE, which has only been implemented for the two-player case, the two users follow identical strategies, which involve randomly deciding whether to wait or transmit, unless they are the only user that has not transmitted, and then they always try to send when not in backoff. In the case of SFCE,  users can employ a shared probabilistic signal to coordinate which user sends next. Initially, this is a uniform choice over the users, but as time progresses the signal favours the users with lower backoff counters as these users have had fewer opportunities to send their packet previously. 

\begin{figure}[t]
\centering
\includegraphics{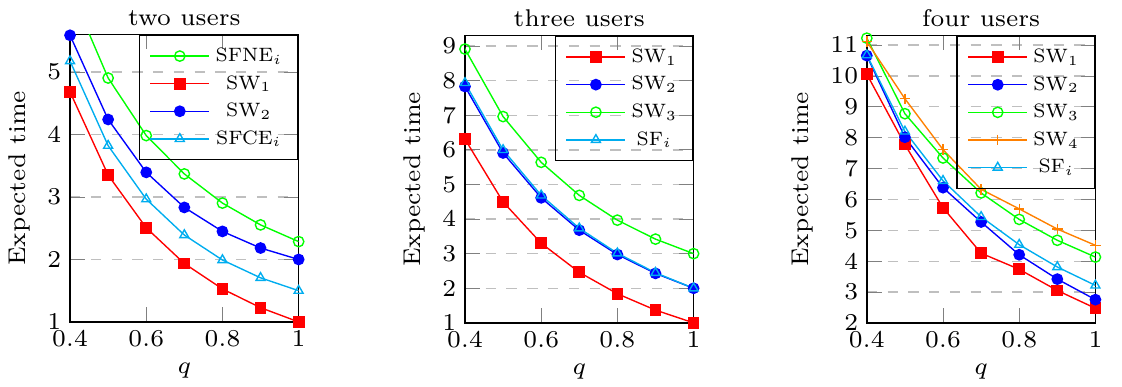}
\vspace*{-0.3cm}
\caption{Aloha: $\nashop{\mathit{usr}_1{:}\cdots{:}\mathit{usr}_m}{\star_1,\star_2}{\min=?}{\rewop{\mathit{time}}{}{\future \mathsf{s}_1}{+}{\cdots}{+}\rewop{\mathit{time}}{}{\future \mathsf{s}_m}}$}
\label{aloha234-fig}
\vspace*{-0.6cm}
\end{figure}

In \figref{aloha234-fig} we have plotted the optimal values for the players, where $\text{SW}_i$ correspond to the optimal values (expected times to send their packets) for player $i$ for both SWNE and SWCE 
for the cases of two, three and four users. We see that the optimal values for the different users under SFNE and SFCE coincide, while under SWNE and SWCE they are different for each user (with the user sending first having the lowest and the user sending last the highest). Comparing the sum of the SWNE (and SWCE) values and that of the SFCE values, we see a small decrease in the sum of less than 2\% of the total, while for SFNE there is a greater difference as the players cannot coordinate, and hence try and send at the same time.

\startpara{Power control} This case study is based on a model of power control in cellular networks from~\cite{BRE13}. In the network there are a number of users that each have a mobile phone. The phones emit signals that the users can strengthen by increasing the phone's power level up to a bound ($\mathit{pow}_{\max}$). A stronger signal can improve transmission quality, but uses more energy and lowers the quality of the transmissions of other phones due to interference. We use the extended model from~\cite{KNPS20}, which adds a probability of failure ($q_\mathit{fail}$) when a power level is increased and assumes each phone has a limited battery capacity ($e_{\max}$). There is a reward structure associated with each phone representing transmission quality, which is dependent on both the phone's power level and the power levels of other phones due to interference. We consider the nonzero-sum property $\nashop{p_1{:}{\cdots}{:}p_m}{\star_1,\star_2}{\max=?}{\rewop{\mathit{r}_1}{}{\future \mathsf{e}_1}{+}{\cdots}{+}\rewop{\mathit{r}_m}{}{\future \mathsf{e}_m}}$, where each user tries to maximise their expected reward before their phone's battery is depleted.

\begin{figure}[t]
\centering
\includegraphics{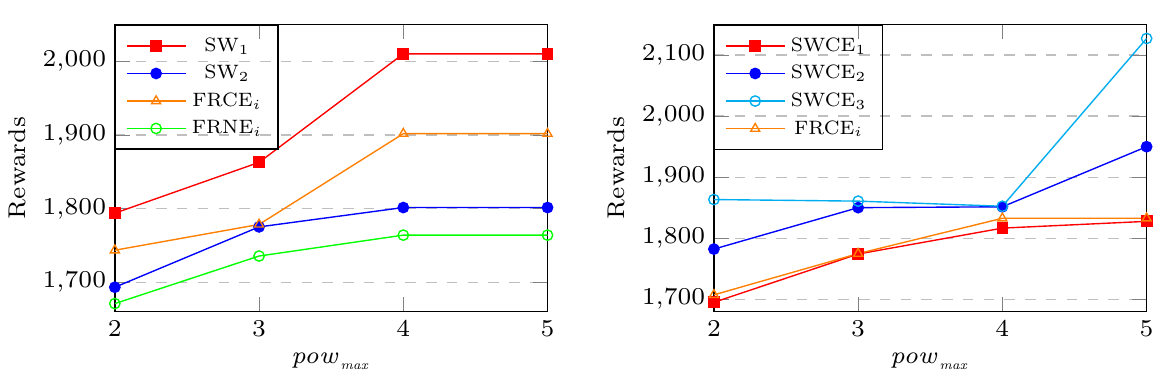}
\vspace*{-0.3cm}
\caption{Power control: $\nashop{p_1{:}{\cdots}{:}p_m}{\star_1,\star_2}{\max=?}{\rewop{\mathit{r}_1}{}{\future \mathsf{e}_1)}{+}{\cdots}{+}\rewop{\mathit{r}_m}{}{\future \mathsf{e}_m}}$}
\label{power-fig}
\vspace*{-0.6cm}
\end{figure}

In \figref{power-fig} we have presented the expected rewards of the players under the synthesised SWCE and SFCE joint strategies. When performing strategy synthesis, in the case of two users the SWNE and SWCE yield the same profile in which, when the users' batteries are almost depleted, one user tries to increase their phone's power level and, if successful, in the next step, the second user then tries to increase their phone's power level. Since the first user's phone battery is depleted when the second tries to increase, this increase does not cause any interference. On the other hand, if the first user fails to increase their power level, then both users increase their battery levels. For the SFCE, the users can coordinate and flip a coin as to which user goes first: as demonstrated by \figref{power-fig} this yields equal rewards for the users, unlike the SWCE. In the case of three users, the SWNE and SWCE differ (we were only able to synthesise SWNE for $\mathit{pow}_{\max}=2$ as for larger values the computation had not completed within the timeout), again users take turns to try and increase their phone's power level. However, here if the users are unsuccessful the SWCE can coordinate as to which user goes next trying to increase their phone's battery level. Through this coordination, the users' rewards can be increased as the battery level of at most one phone increases at a time, which limits interference. On the other hand, for the SWNE users must decide independently whether to increase their phone's battery level and they each randomly decide whether to do so 
or not.

\startpara{Public good} We next consider a variant of a \emph{public good} game~\cite{HHCN19}, based on the one presented in \cite{KNPS20} for the two-player case. In this game a number of players each receive an initial amount of capital ($e_\mathit{init}$) and, in each of $r_\mathit{max}$ months, can invest none, half or all of their current capital. The total invested by the players in a month is multiplied by a factor $f$ and distributed equally among the players before the start of the next month. The aim of the players is to maximise their expected capital which is represented by the formula: $\nashop{p_1{:}\cdots{:}p_m}{\star_1,\star_2}{\max=?}{\rewop{c_1}{}{\sinstant{=r_\mathit{max}}}{+}{\cdots}{+}\rewop{c_m}{}{\sinstant{=r_\mathit{max}}}}$.

\begin{figure}[t]
\centering
\includegraphics{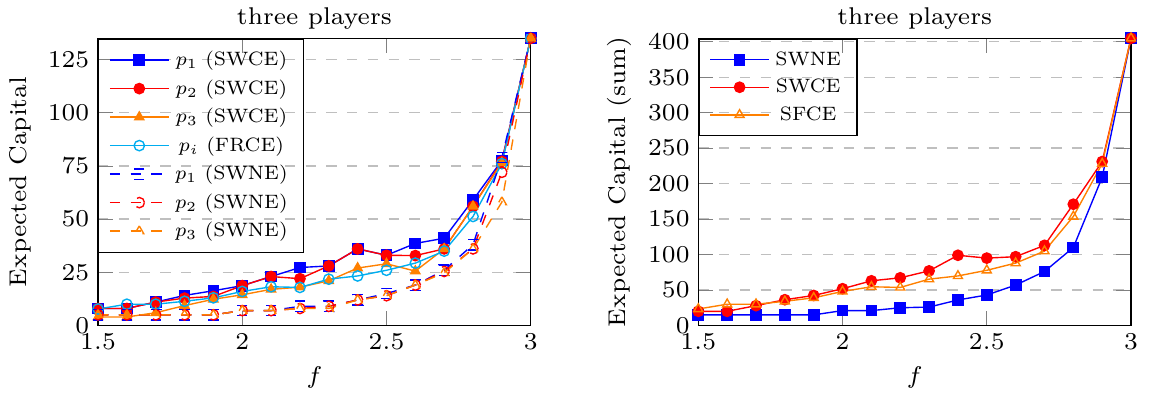}
\vspace*{-0.3cm}
\caption{Public good: $\nashop{p_1{:}\cdots{:}p_m}{\star_1,\star_2}{\max=?}{\rewop{c_1}{}{\sinstant{=r_\mathit{max}}}{+}{\cdots}{+}\rewop{c_m}{}{\sinstant{=r_\mathit{max}}}}$}\label{fig:pgg}
\vspace*{-0.6cm}
\end{figure}

\figref{fig:pgg} plots, for the three-player model, both the expected capital of individual players and the total expected capital after three months for the SWNE, SWCE and SFNE as the parameter $f$ varies. As the results demonstrate the players benefit, both as individuals and as a population, by coordinating through a correlated strategy. In addition, under the SFCE, all players receive the same expected capital with only a small decrease in the sum from that of the SWCE. 

\startpara{Investors} The final case study concerns a concurrent multi-player version of futures market investor model of~\cite{MM07}, in which a number of investors (the players) interact with a probabilistic stock market.
In successive months, the investors choose whether to invest, wait or cash in their shares, while at the same time the market decides with probability $\mathit{pbar}$ to bar each investor, with the restriction that an investor cannot be barred two months in a row or in the first month, and then the values of shares and cap on values are updated probabilistically.

\begin{figure}[t]
\centering
\includegraphics{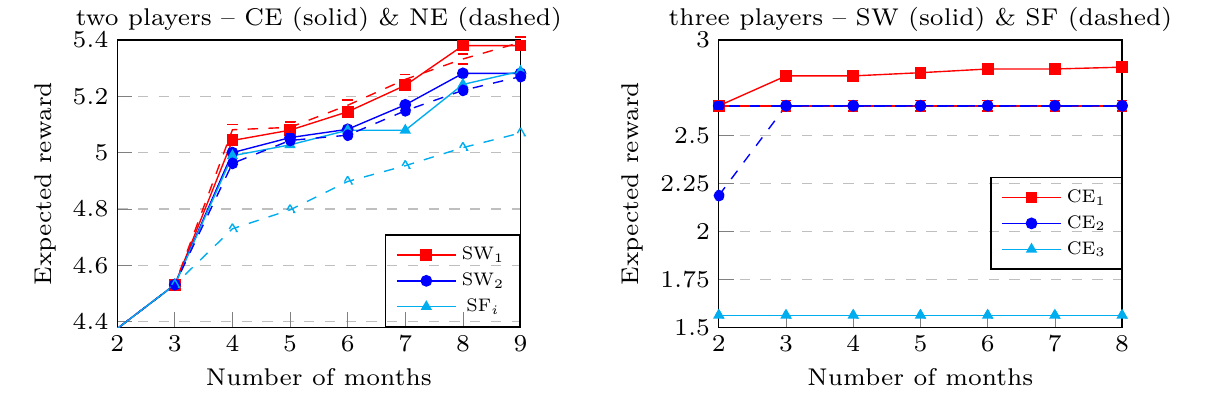}
\vspace*{-0.6cm}
\caption{Investors: $\nashop{\mathit{inv}_1{:}{\cdots}{:}\mathit{inv}_m}{\star_1,\star_2}{\max=?}{\rewop{\mathit{pf}_1}{}{\future \mathsf{cin}_1}{+}{\cdots}{+}\rewop{\mathit{pf}_m}{}{\future \mathsf{cin}_m}}$}\label{fig:inv}
\vspace*{-0.6cm}
\end{figure}

We consider both two- and three-player models, where each investor tries to maximise its individual profit represented by the following nonzero-sum property:
$\nashop{\mathit{inv}_1{:}{\cdots}{:}\mathit{inv}_m}{\star_1,\star_2}{\max=?}{\rewop{\mathit{pf}_1}{}{\future \mathsf{cin}_1}{+}{\cdots}{+}\rewop{\mathit{pf}_m}{}{\future \mathsf{cin}_m}}$. In \figref{fig:inv} we have plotted the different optimal values for NE and CE of the two-player game and the different optimal values for CE of the three-player game (the computation of NE values timed out for the three player case). As the results demonstrate, again we see that the coordination that CEs offer can improve the returns of the players and that, although considering social fairness does decrease the returns of some players, this is limited, particularly for CEs.

\section{Conclusions}\label{conc-sect}

We have presented novel techniques for game-theoretic verification
of probabilistic multi-agent systems, focusing on correlated equilibria
and a notion of social fairness. We began with the simpler case of normal form
games and then extended this to concurrent stochastic games, and used temporal logic
to formally specify equilibria. We proposed algorithms for equilibrium synthesis,
implemented them and illustrated their benefits, in terms of efficiency and fairness,
on case studies from a range of application domains.

Future work includes exploring the use of further game-theoretic topics
within this area, such as techniques for mechanism design
or other concepts such as Stackelberg equilibria. We plan to implement SFCE computation in Gurobi using the \emph{big-M method}~\cite{GNS90} to encode implications and techniques from~\cite{SP17} to encode conjunctions, which should yield a significant speed-up in their computation.

\startpara{Acknowledgements}
This project was funded by the ERC under the European Union’s Horizon 2020 research and innovation programme (\href{http://www.fun2model.org}{FUN2MODEL}, grant agreement No.~834115).

\bibliographystyle{splncs04.bst}
\bibliography{bib}

\newpage
\appendix
\section{Model Checking Algorithm for Nonzero-Sum Formulae}\label{appendix}

In this appendix we give the complete model checking algorithm for verifying a formula of the form $\phi = \nashop{C_1{:}{\cdots}:C_m}{\star_1,\star_2}{\opt\sim x}{\theta}$ on CSG $\game$ where $\opt=\max$,
i.e., computing the vector of values $\V_{\game^\cC}(s,\star_1,\star_2,\theta)$
for the $m$-player coalition game $\V_{\game^\cC}$.
As stated in \sectref{csg-sect}, the case when $\opt = \min$ can be treated by negating all utilities and maximising. We first give the cases when all the objectives in $\theta$ are of the same type and then when they are mixed but all finite-horizon objectives. Finally, when there is a combination of finite- and infinite-horizon objectives, we show how the game under study and $\theta$ is modified, which includes making all objectives infinite-horizon.

\startpara{Next} If $\theta = \probop{}{\next \phi^1}{+}\cdots{+}\probop{}{\next \phi^m}$, then 
$\V_{\game^C}(s,\star_1,\star_2,\theta)$ equals optimal values of the game $\nfgame = (M,A^\cC,u)$ with respect to the criterion $\star_2$ and of equilibria type $\star_1$ in which for any $1\leq l\leq m$ and $a \in A^\cC$: 
\[
u^l(a) = \mbox{$\sum_{s' \in \Sat(\phi^l)}$} \, \delta^\cC(s,a)(s')
\]
for $s' \in S$.

\startpara{Bounded Probabilistic Until}
If $\theta = \probop{}{\phi^1_1 \ \untilop^{\leq k_1} \ \phi^1_2}{+}\cdots{+}\probop{}{\phi^{m}_1 \ \untilop^{\leq k_m} \ \phi^{m}_2}$, we compute optimal values of the objectives for the nonzero-sum formulae:
\[
\theta_n = \probop{}{\phi^1_1 \ \untilop^{\leq k_1-n}\ \phi^1_2}{+}\cdots{+}\probop{}{\phi^{m}_1 \ \untilop^{\leq k_m-n} \ \phi^{m}_2}
\]
for $0\leq n \leq k$ recursively, where $k=\max\{k_1,\dots,k_l\}$ and $\V_{\game^\cC}(s,\star_1,\star_2,\theta)=\V_{\game^\cC}(s,\star_1,\star_2,\emptyset,\emptyset,\theta_0)$. For any state $s$ and $0\leq n \leq k$, $D,E \subseteq M$ such that $D \cap E = \emptyset$ we have:
\[
\V_{\game^\cC}(s,\star_1,\star_2,D,E,\theta_n) = \left\{ \begin{array}{cl}
(\eta_{D}(1),\dots,\eta_{D}(m)) & \; \mbox{if $D \cup E = M$} \\
\V_{\game^\cC}(s,\star_1,\star_2,D \cup D',E,\theta_n) & \; \mbox{else if $D' \neq \varnothing$} \\
\V_{\game^\cC}(s,\star_1,\star_2,D,E \cup E',\theta_n) & \; \mbox{else if $E' \neq \varnothing$} \\
\val(\nfgame,\star_1,\star_2) & \; \mbox{otherwise}
\end{array} \right. 
\]
where 
\begin{align*}
D' & = \;  \{1 \leq l \leq m {\setminus} (D\cup E) \mid s \in \Sat(\phi^l_2)\} \\
E' & = \;  \{1 \leq l \leq m {\setminus} (D\cup E) \mid s \in \Sat(\neg \phi^l_1 \wedge \neg \phi^l_2)\}
\end{align*}
and $\val(\nfgame,\star_1,\star_2)$ equals optimal values of the game $\nfgame = (M,A^\cC,u)$ with respect to the criterion $\star_2$ and of equilibria type $\star_1$ in which for any $1 \leq l \leq m$ and $a \in A^\cC$:
\[ 
u_l(a) = \left\{
\begin{array}{cl}
1 & \;\mbox{if $l \in D$} \\
0 & \;\mbox{else if $l \in E$} \\
0 & \;\mbox{else if $n_l-n \leq 0$} \\
\sum_{s' \in S} \delta^{\cC}(s,a)(s') \cdot v^{s',l}_{n-1} & \;\mbox{otherwise}
\end{array} \right.
\]
and $(v^{s',1}_{n-1},v^{s',2}_{n-1},\dots,v^{s',m}_{n-1}) = \V_{\game^\cC}(s',\star_1,\star_2,D,E,\theta_{n-1})$ for $s' \in S$.
\startpara{Instantaneous Rewards}
If $\theta = \rewop{r_1}{}{\sinstant{= k_1}}+\cdots+\rewop{r_{m}}{}{\sinstant{= k_{m}}}$, we compute optimal values of the objectives for the nonzero-sum formulae:
\[
\theta_{n}=\rewop{r_1}{}{\sinstant{=n_1-n}}+\dots+\rewop{r_{m}}{}{\sinstant{=n_l-n}}
\]
for $0\leq n \leq k$ recursively, where $k=\max\{k_1,\dots,k_l\}$ and $\V_{\game^\cC}(s,\star_1,\star_2,\theta) = \V_{\game^\cC}(s,\star_1,\star_2, \theta_0)$. For any state $s$ and $0\leq n \leq k$, $\V_{\game^\cC}(s,\star_1,\star_2,\theta_{n})$ equals optimal values of the game $\nfgame = (M,A^{\cC},u)$ with respect to the criterion $\star_2$ and of equilibria type $\star_1$ in which for any $1 \leq l \leq m$ and $a \in A^{\cC}$:
\[ 
u_l(a) = \left\{
\begin{array}{cl}
0  & \;\mbox{if $n_l-n < 0$} \\
\sum_{s' \in S} \delta^{\cC}(s,a)(s') \cdot r^l_S(s')  & \;\mbox{else if $n_l-n = 0$} \\
\sum_{s' \in S} \delta^{\cC}(s,a)(s') \cdot v^{s',l}_{n+1} & \;\mbox{otherwise}
\end{array} \right.
\]
and $(v^{s',1}_{n+1},\dots,v^{s',m}_{n+1}) = \V_{\game^\cC}(s',\star_1,\star_2,\theta_{n+1})$ for $s' \in S$.
\startpara{Bounded Cumulative Rewards}
If $\theta = \rewop{r_1}{}{\scumul{\leq k_1}}+\cdots+\rewop{r_{m}}{}{\scumul{\leq k_{m}}}$, we compute optimal values of the objectives for the nonzero-sum formulae:
\[
\theta_{n}=\rewop{r_1}{}{\scumul{\leq n_1-n}}+\dots+\rewop{r_l}{}{\scumul{\leq n_{m}-n}}
\]
for $0\leq n \leq k$ recursively, where $k=\max\{k_1,\dots,k_l\}$ and $\V_{\game^\cC}(s,\star_1,\star_2,\theta)$ $= \V_{\game^\cC}(s,\star_1,\star_2, \theta_0)$. For any state $s$ and $0\leq n \leq k$,
$\V_{\game^\cC}(s,\star_1,\star_2,\theta_n)$ equals optimal values of the game $\nfgame = (M,A^{\cC},u)$ with respect to the criterion $\star_2$ and of equilibria type $\star_1$ in which for any $1 \leq l \leq m$ and $a \in A^{\cC}$:
\[ 
u_l(a) = \left\{
\begin{array}{cl}
0 & \;\mbox{if $n_l-n \leq 0$} \\
r^l_S(s) + r^l_A(s,a) + \sum_{s' \in S} \delta^{\cC}(s,a)(s') \cdot v^{s',l}_{n+1} & \;\mbox{otherwise}
\end{array} \right.
\]
and $(v^{s',1}_{n+1},\dots,v^{s',m}_{n+1}) = \V_{\game^\cC}(s',\star_1,\star_2,\theta_{n+1})$ for $s' \in S$.
\startpara{Probabilistic Until}
If $\theta = \probop{}{\phi^1_1 \until \phi^1_2}{+}\cdots{+}\probop{}{\phi^{m}_1 \until \phi^{m}_2}$, the optimal values can be computed through value iteration as the limit $\V_{\game^\cC}(s,\star_1,\star_2,\theta) = \lim_{n \ra \infty} \V_{\game^\cC}(s,\star_1,\star_2,\theta,n)$ where $\V_{\game^\cC}(s,\star_1,\star_2,\theta,n)=\V_{\game^\cC}(s,\star_1,\star_2,\emptyset,\emptyset,\theta,n)$ and for any $D,E \subseteq M$ such that $D \cap E = \emptyset$:
\[
\V_{\game^\cC}(s,\star_1,\star_2,D,E,\theta,n) = \left\{ \begin{array}{cl}
(\eta_{D}(1),\dots,\eta_{D}(m)) & \; \mbox{if $D \cup E = M$} \\
(\eta_{\Sat(\phi^1_2)}(s),\dots,\eta_{\Sat(\phi^{m}_2)}(s)) & \; \mbox{else if $n=0$} \\
\V_{\game^\cC}(s,\star_1,\star_2,D \cup D',E,\theta,n) & \; \mbox{else if $D' \neq \varnothing$} \\
\V_{\game^\cC}(s,\star_1,\star_2,D,E \cup E',\theta,n) & \; \mbox{else if $E' \neq \varnothing$} \\
\val(\nfgame,\star_1,\star_2) & \; \mbox{otherwise}
\end{array} \right. 
\]
where:
\begin{align*}
D' & = \; \{1 \leq l \leq m {\setminus} (D\cup E) \mid s \in \Sat(\phi^l_2)\} \\
E' & = \;  \{1 \leq l \leq m {\setminus} (D\cup E) \mid s \in \Sat(\neg \phi^l_1 \wedge \neg \phi^l_2)\}
\end{align*}
and $\val(\nfgame,\star_1,\star_2)$ equals optimal values of the game $\nfgame = (M,A^\cC,u)$ with respect to the criterion $\star_2$ and of equilibria type $\star_1$ in which for any $1 \leq l \leq m$ and $a \in A^\cC$:
\[ 
u_l(a) = \left\{
\begin{array}{cl}
1 & \;\mbox{if $l \in D$} \\
0 & \;\mbox{else if $l \in E$} \\
\sum_{s' \in S} \delta^{\cC}(s,a)(s') \cdot v^{s',l}_{n-1} & \;\mbox{otherwise}
\end{array} \right.
\]
and $(v^{s',1}_{n-1},v^{s',2}_{n-1},\dots,v^{s',m}_{n-1}) = \V_{\game^\cC}(s',\star_1,\star_2,D,E,\theta,n-1)$ for $s' \in S$.
\startpara{Expected Reachability} If $\theta = \rewop{r_1}{}{\future \ap^1}{+}\cdots{+}\rewop{r_{m}}{}{\future \ap^{m}}$, the optimal values can be computed through value iteration as the limit 
$\V_{\game^\cC}(s,\star_1,\star_2,\theta) = \lim_{n \ra \infty} \V_{\game^\cC}(s,\star_1,\star_2,\theta,n)$ where $\V_{\game^\cC}(s,\star_1,\star_2,\theta,n)=\V_{\game^\cC}(s,\star_1,\star_2,\emptyset,\theta,n)$ and for any $D \subseteq M$:
\[
\V_{\game^\cC}(s,\star_1,\star_2,D,\theta,n) = \left\{ \begin{array}{cl}
\mathbf{0}_{m} & \; \mbox{if $D = M$} \\
\mathbf{0}_{m} & \; \mbox{else if $n= 0$} \\
\V_{\game^\cC}(s,\star_1,\star_2,D \cup D',\theta,n) & \; \mbox{else if $D' \neq \varnothing$} \\
\val(\nfgame,\star_1,\star_2) & \; \mbox{otherwise}
\end{array} \right. 
\]
$D' = \{1 \leq l \leq m {\setminus} D \mid \ap^l \in L(s)  \}$ and $\val(\nfgame,\star_1,\star_2)$ equals optimal values of the NFG $\nfgame = (M,A^\cC,u)$ with respect to the criterion $\star_2$ and of equilibria type $\star_1$ in which for any $1 \leq l \leq m$ and $a \in A^\cC$:
\[ 
u_l(a) = \left\{
\begin{array}{cl}
0 & \;\mbox{if $l \in D$} \\
r^l_S(s) + r^l_A(s,a) + \sum_{s' \in S} \delta^{\cC}(s,a)(s') \cdot v^{s',l}_{n-1} & \;\mbox{otherwise}
\end{array} \right.
\]
and $(v^{s',1}_{n-1},v^{s',2}_{n-1},\dots,v^{s',m}_{n-1}) = \V_{\game^\cC}(s',\star_1,\star_2,D,\theta,n-1)$ for $s' \in S$.

\startpara{Mixed Finite-Horizon Probabilistic Objectives}
If $\theta = \probop{}{\psi_1}{+}\cdots{+}\probop{}{\psi_m}$ where $\psi_l$ is either of the form $\next \phi^l$ or $\phi^l_1 \ \untilop^{\leq k_l} \ \phi^l_2$ for $1 \leq l \leq m$, we compute optimal values of the objectives for the nonzero-sum formulae $\theta_n = \probop{}{\psi^n_1}{+}\cdots{+}\probop{}{\psi^n_m}$ for $0\leq n \leq k$ recursively, where:
\[
\psi^n_l = \left\{ \begin{array}{cl}
\next \phi^l & \mbox{if $\psi_l = \next \phi^l$} \\
\phi^{m}_1 \ \untilop^{\leq k_m-n} \ \phi^{m}_2 & \mbox{otherwise.}
\end{array} \right.
\]
$k=\max\{k_1,\dots,k_l\}$ and $\V_{\game^\cC}(s,\star_1,\star_2,\theta)=\V_{\game^\cC}(s,\star_1,\star_2,\emptyset,\emptyset,\theta_0)$. Letting $M_\next = \{ 1 \leq l \leq m \mid \psi_l = \next \phi^l \}$, for any state $s$ and $0\leq n \leq k$, $D,E \subseteq M$ such that $D \cap E = \emptyset$:
\[
\V_{\game^\cC}(s,\star_1,\star_2,D,E,\theta_n) = \left\{ \begin{array}{cl}
(\eta_{D}(1),\dots,\eta_{D}(m)) & \; \mbox{if $D \cup E = M$} \\
\V_{\game^\cC}(s,\star_1,\star_2,D \cup D',E,\theta_n) & \; \mbox{else if $D' \neq \varnothing$} \\
\V_{\game^\cC}(s,\star_1,\star_2,D,E \cup E',\theta_n) & \; \mbox{else if $E' \neq \varnothing$} \\
\val(\nfgame,\star_1,\star_2) & \; \mbox{otherwise}
\end{array} \right. 
\]
where if $k=1$, then:
\begin{align*}
D' &= \;  \{1 \leq l \leq m {\setminus} (D\cup E \cup M_\next) \mid s \in \Sat(\phi^l_2)\} \cup \{ 1 \leq l \leq m_\next \mid s \in \Sat(\phi^l) \} \\
E' &= \; \{1 \leq l \leq m {\setminus} (D\cup E \cup M_\next) \mid s \in \Sat(\neg \phi^l_1 \wedge \neg \phi^l_2)\} \cup \{ 1 \leq l \leq m_\next \mid s \not\in \Sat(\phi^l) \} 
\end{align*}
and otherwise:
\begin{align*}
D' &= \;  \{1 \leq l \leq m {\setminus} (D\cup E \cup M_\next) \mid s \in \Sat(\phi^l_2)\} \\
E' &= \; \{1 \leq l \leq m {\setminus} (D\cup E \cup M_\next) \mid s \in \Sat(\neg \phi^l_1 \wedge \neg \phi^l_2)\}
\end{align*}
and $\val(\nfgame,\star_1,\star_2)$ equals optimal values of the game $\nfgame = (M,A^\cC,u)$ with respect to the criterion $\star_2$ and of equilibria type $\star_1$ in which for any $1 \leq l \leq m$ and $a \in A^\cC$:
\[ 
u_l(a) = \left\{
\begin{array}{cl}
1 & \;\mbox{if $l \in D$} \\
0 & \;\mbox{else if $l \in E$} \\
\sum_{s' \in S} \delta^{\cC}(s,a)(s') \cdot v^{s',l}_{n-1} & \;\mbox{otherwise}
\end{array} \right.
\]
and $(v^{s',1}_{n-1},v^{s',2}_{n-1},\dots,v^{s',m}_{n-1}) = \V_{\game^\cC}(s',\star_1,\star_2,D,E,\theta,n-1)$ for $s' \in S$.

\startpara{Mixed Finite-Horizon Reward Objectives}
If $\theta = \rewop{r_1}{}{\rho_1+\cdots+\rewop{r_{m}}{}{\rho^l}}$ and $\rho_l$ is either of the form $\sinstant{= k_l}$ or $\scumul{\leq k_1}$ for $1 \leq l \leq m$, we compute optimal values of the objectives for the nonzero-sum formulae $\theta_{n}=\rewop{r_1}{}{\rho_1^n+\cdots+\rewop{r_{m}}{}{\rho_m^n}}$ where:
\[
\rho^n_l = \left\{ \begin{array}{cl}
\sinstant{= k_l-n} & \mbox{if $\rho= \sinstant{= k_l}$} \\
\scumul{\leq n_1-n} & \mbox{otherwise.}
\end{array} \right.
\]
for $0\leq n \leq k$ recursively, where $k=\max\{k_1,\dots,k_l\}$ and $\V_{\game^\cC}(s,\star_1,\star_2,\theta)$ $= \V_{\game^\cC}(s,\star_1,\star_2, \theta_0)$.  Letting $M_\mathtt{I} = \{ 1 \leq l \leq m \mid \rho_l = \sinstant{= k_l} \}$,  any state $s$ and $0\leq n \leq k$,
$\V_{\game^\cC}(s,\star_1,\star_2,\theta_n)$ equals optimal values of the game $\nfgame = (M,A^{\cC},u)$ with respect to the criterion $\star_2$ and of equilibria type $\star_1$ in which for any $1 \leq l \leq m$ and $a \in A^{\cC}$:
\[ 
u_l(a) = \left\{
\begin{array}{cl}
0 & \; \mbox{if $n_l-n < 0 \wedge l \in M_\mathtt{I}$} \\
\sum_{s' \in S} \delta^{\cC}(s,a)(s') \cdot r^l_S(s')  & \; \mbox{else if $n_l{-}n = 0 \wedge l \in M_\mathtt{I}$} \\
\sum_{s' \in S} \delta^{\cC}(s,a)(s') \cdot v^{s',l}_{n+1} & \; \mbox{else if $l \in M_\mathtt{I}$} \\
0 & \; \mbox{else if $n_l{-}n \leq 0$} \\
r^l_S(s) + r^l_A(s,a) + \sum_{s' \in S} \delta^{\cC}(s,a)(s') \cdot v^{s',l}_{n+1} & \; \mbox{otherwise}
\end{array} \right.
\]
and $(v^{s',1}_{n+1},\dots,v^{s',m}_{n+1}) = \V_{\game^\cC}(s',\star_1,\star_2,\theta_{n+1})$ for $s' \in S$.

\startpara{Mixed Finite- and Infinite-Horizon Objectives}
In the case when $\theta$ is a mixture of finite- and infinite-horizon objectives we reduce the problem to finding values for a sum of modified (infinite-horizon) objectives $\theta'$ on a modified game $\game'$. This approach is based on the standard construction for converting the verification of finite-horizon properties to infinite-horizon properties~\cite{Put94}. The modified game has states of the form $(s,n)$, where $s$ is a state of $\game^C$, $n \in \Nset$ and the optimal values $\V_{\game^C}(s,\theta)$ are given by the optimal values $\V_{\game'}((s,0),\theta')$. Therefore, since we require the SWNE values for states of the original game, in the modified game the set of initial states equals $\{ (s,0) \mid s \in S \}$. We let $k_\theta$ equal the maximum bound appearing in objectives (if $\theta$ only contains next and unbounded until formulae, then we let $k_\theta=1$). We first construct the game $\game' = (M, S', \bar{S}', A^C, \Delta', \delta', \AP' , \lab')$ where the states, action assignment functions and probabilistic transition function are defined as follows.
\begin{itemize}
 \setlength{\itemsep}{2pt}
  \setlength{\parskip}{2pt}
  \setlength{\parsep}{2pt}
\item
$S' = \{ (s,n) \mid s \in S \wedge 0 \leq n \leq k_\theta+1 \}$ and $\bar{S}' = \{ (s,0) \mid s \in S \}$.
\item $\Delta'((s,n)) = \Delta^C(s)$ for $(s,n) \in S'$.
\item For any $(s,n),(s',n') \in S'$ and $a \in A^C$:
\[
\delta'((s,n),a)((s',n')) = \begin{cases}
\delta^C(s,a)(s') & \mbox{if $0 \leq n \leq 1$ and $n' = n{+}1$} \\
\delta^C(s,a)(s') & \mbox{else if $n = n' = k_\theta+1$} \\
0 & \mbox{otherwise.}
\end{cases}
\]
\end{itemize}
It remains to define the atomic propositions and labelling function of the game $\game'$ and new objectives and reward structures in the case of reward objectives. In the case when $\theta$ is of the form $\probop{}{\psi_1}{+}\cdots{+}\probop{}{\psi_m}$, we set $\AP' = \cup_{l=1}^m \AP_{\! \rho_l}$ and $\theta'=\probop{}{\psi_1'}{+}\cdots{+}\probop{}{\psi_m'}$ where:
\begin{itemize}
 \setlength{\itemsep}{2pt}
  \setlength{\parskip}{2pt}
  \setlength{\parsep}{2pt}
\item 
if $\psi_l= \next \phi^l$, then 
$\AP_{\psi_l} = \{ \ap_{\phi^l} \}$,
$\psi_l' = \future \ap_{\phi^l}$ and for any $(s,n) \in S'$ we have
$\ap_{\phi^l} \in \lab'((s,n))$ if and only if $s \in \Sat(\phi^l)$ and $n = 1$;
\item 
if $\psi_l= \phi_1^l \buntilp{k_l} \phi_2^l$, then 
$\AP_{\psi_l} = \{ \ap_{\phi_2^l},\ap_{\phi_2^l} \}$, 
$\psi_l' =\ap_{\phi_1^l} \until \ap_{\phi_2^l}$ and for any $(s,n) \in S'$ and $1 \leq j \leq 2$ we have
$\ap_{\phi_j^l} \in \lab'((s,n))$ if and only if $s \in \Sat(\phi_j^l)$ and $n \leq k_l$;
\item
if $\psi_l= \phi_1^l \until \phi_2^l$, then 
$\AP_{\psi_l} = \{ \ap_{\phi_2^l},\ap_{\phi_2^l} \}$,
$\psi_l' = \ap_{\phi_1^l} \until \ap_{\phi_2^l}$ and for any $(s,n) \in S'$ and $1 \leq j \leq 2$ we have
$\ap_{\phi_j^2} \in \lab'((s,n))$ if and only if $s \in \Sat(\phi_j^2)$.
\end{itemize}
On the other hand, when $\theta$ is of the form $\rewop{r_1}{}{\rho_1}+\cdots+\rewop{r_{m}}{}{\rho_l}$, we set $\AP' = \cup_{l=1}^m \AP_{\! \rho_l}$ and $\theta'= \rewop{r_1'}{}{\rho_1'}+\cdots+\rewop{r_{m}'}{}{\rho_l'}$ where:
\begin{itemize}
 \setlength{\itemsep}{2pt}
  \setlength{\parskip}{2pt}
  \setlength{\parsep}{2pt}
\item
if $\rho_l=\sinstant{=k_l}$, then $\AP_{\! \rho_l} = \{ \ap_{k_l+1} \}$, $r_l' =(r'_{l,A},r'_{l,S})$, $\rho_l' = \future \ap_{k_l+1}$ and for any $(s,n) \in S'$ and $a \in A$ we have
\begin{itemize}
\item[--]
$\ap_{k_l+1} \in \lab'((s,n))$ if and only if $n = k_l{+}1$;
\item[--]
$r'_{l,A}((s,n),a)=0$;
\item[--]
$r'_{l,S}((s,n))=r^l_{S}(s)$ if $n = k_l$ and equals $0$ otherwise;
\end{itemize}
\item
if $\rho_l=\scumul{\leq k_l}$, then $\AP_{\! \rho_l} = \{ \ap_{k_l} \}$, $r_l' =(r'_{l,A},r'_{l,S})$, $\rho_l' = \future \ap_{k_l}$ and for any $(s,n) \in S'$ and $a \in A$ we have
\begin{itemize}
\item[--]
$\ap_{k_l} \in \lab'((s,n))$ if and only if $n = k_l$;
\item[--]
$r'_{l,A}((s,n),a)=r^l_{A}(s,a)$ if $n \leq k_l-1$ and equals 0 otherwise;
\item[--]
$r'_{l,S}((s,n))=r^l_{S}(s)$ if $n \leq k_l-1$ and equals 0 otherwise;
\end{itemize}
\item if $\rho_l=\future \phi^l$, then $\AP_{\! \rho_l} = \{ \ap_{\phi^l} \}$, $r_l' =(r'_{l,A},r'_{l,S})$, $\rho_l' = \future \ap_{\phi^l}$ and for any $(s,n) \in S'$ and $a \in A$ we have
\begin{itemize}
\item[--]
$\ap_{\phi^l} \in \lab'((s,n))$ if and only if $s \in \Sat(\phi^l)$;
\item[--]
$r'_{l,A}((s,n),a)=r^l_{A}(s)(a)$ and $r'_{l,S}((s,n))=r^l_{S}(s)$.
\end{itemize}
\end{itemize}

\end{document}

%% file: macros-dave.tex
\DeclareMathOperator{\opt}{opt}
\renewcommand{\emptyset}{\varnothing}

\usepackage{scalerel}

\newcommand{\appref}[1]{Appendix~\ref{#1}}
\newcommand{\sectref}[1]{Section~\ref{#1}}

\newcommand{\figref}[1]{Figure~\ref{#1}}
\newcommand{\tabref}[1]{Table~\ref{#1}}

\newcommand{\eqnref}[1]{(\ref{#1})}

\newcommand{\defref}[1]{Definition~\ref{#1}}

\newcommand{\defdefref}[2]{Definitions~\ref{#1} and \ref{#2}}

\spnewtheorem{assumption}{Assumption}{\bfseries}{\itshape}
\newcounter{exampcount}
\newenvironment{examp}
{\refstepcounter{exampcount}
\vskip6pt\noindent
{\bf Example \arabic{exampcount}.}}
{\hfill$\blacksquare$\vskip6pt}

\newcommand{\startpara}[1]{{%
\vskip6pt\noindent
{\bf #1.}}}

\def\ra{{\rightarrow}}

\def\cC{{\mathcal{C}}}

\def\Nset{\mathbb{N}}
\def\Qset{\mathbb{Q}}
\def\Rset{\mathbb{R}}

\def\Eset{\mathbb{E}}

\def\ra{\rightarrow} 
\def\rmdef{\,\stackrel{\mbox{\rm {\tiny def}}}{=}}
\newcommand{\true}{\mathtt{true}} 

\renewcommand{\leq}{\leqslant}
\renewcommand{\geq}{\geqslant}

\def\squareforqed{\hbox{\rlap{$\sqcap$}$\sqcup$}}
\def\qed{\ifmmode\squareforqed\else{\unskip\nobreak\hfil
\penalty50\hskip1em\null\nobreak\hfil\squareforqed
\parfillskip=0pt\finalhyphendemerits=0\endgraf}\fi}

\newcommand\game{{\sf G}}
\newcommand\nfgame{{\sf N}}

\newcommand\sinit{{\bar{s}}}

\newcommand\dist{{\mathit{Dist}}}

\newcommand\Prob{{\mathit{Prob}}}

\newcommand\val{{\mathit{val}}}

\newcommand{\last}{\mathit{last}}
\newcommand{\ipaths}{\mathit{IPaths}}
\newcommand{\fpaths}{\mathit{FPaths}}

\def\AP{{\mathit{AP}}}

\def\Sat{{\mathit{Sat}}}
\newcommand{\coalition}[1]{\langle \! \langle {#1} \rangle \! \rangle}

\def\next{{X\,}}
\def\until{{\ {\cal U}\ }}
\def\buntil{{\ {\cal U}^{\leq k}\ }}

\def\next{{\mathtt X\,}}
\def\until{{\ \mathtt{U}\ }}
\def\untilop{{\mathtt{U}}}

\def\buntil{{\ \mathtt{U}^{\leq k}\ }}

\newcommand{\buntilp}[1]{{\ \mathtt{U}^{\leq #1}\, }}

\def\future{{\mathtt{F}\ }}

\newcommand{\scumul}[1]{\mathtt{C}^{#1}} 

\newcommand{\ap}{\mathsf{a}}
\newcommand{\sinstant}[1]{\mathtt{I}^{#1}} 

\newcommand\V{{\mathtt V}}
\newcommand\probopP{{\mathtt P}}
\newcommand\nashop[4]{\coalition{#1}({#2})_{#3}(#4)}

\newcommand\NE{\mbox{\sc ne}}
\newcommand\CE{\mbox{\sc ce}}
\newcommand\SW{\mbox{\sc sw}}
\newcommand\SF{\mbox{\sc sf}}

\newcommand\probop[2]{\probopP_{#1}[\,{#2}\,]}

\newcommand\rewopR{{\mathtt R}}
\newcommand\rewop[3]{\rewopR^{#1}_{#2}[\,{#3}\,]}

\newcommand{\lab}{{\mathit{L}}}




%% file: main.bbl
\begin{thebibliography}{10}
\providecommand{\url}[1]{\texttt{#1}}
\providecommand{\urlprefix}{URL }
\providecommand{\doi}[1]{https://doi.org/#1}

\bibitem{dA97a}
de~Alfaro, L.: Formal Verification of Probabilistic Systems. Ph.D. thesis,
  Stanford University (1997)

\bibitem{AKMMR19}
Aminof, B., Kwiatkowska, M., B.~Maubert, B., Murano, A., Rubin, S.:
  Probabilistic strategy logic. In: Proc.\ {IJCAI}'19. pp. 32--38 (2019)

\bibitem{Aum74}
Aumann, R.: Subjectivity and correlation in randomized strategies. Journal of
  Mathematical Economics  \textbf{1}(1),  67--96 (1974)

\bibitem{BK08}
Baier, C., Katoen, J.P.: Principles of Model Checking. MIT Press (2008)

\bibitem{BK98}
Baier, C., Kwiatkowska, M.: Model checking for a probabilistic branching time
  logic with fairness. Distributed Computing  \textbf{11}(3),  125--155 (1998)

\bibitem{BMMSS21}
Banerjee, T., Majumdar, R., Mallik, K., Schmuck, A.K., Soudjani, S.: Fast
  symbolic algorithms for omega-regular games under strong transition fairness.
  Tech. Rep. MPI-SWS-2020-007r, Max Planck Institute (2021)

\bibitem{BRE13}
Brenguier, R.: {PRALINE}: A tool for computing {N}ash equilibria in concurrent
  games. In: Sharygina, N., Veith, H. (eds.) Proc.\ {CAV}'13. LNCS, vol.~8044,
  pp. 890--895. Springer (2013),
  \href{http://www.lsv.fr/Software/praline/}{lsv.fr/Software/praline/}

\bibitem{CF11}
Chatterjee, K., Fijalkow, N.: A reduction from parity games to simple
  stochastic games. EPTCS  \textbf{54},  74--86 (2011)

\bibitem{CH08}
Chatterjee, K., Henzinger, T.: Value iteration. In: 25 Years of Model Checking.
  LNCS, vol.~5000, pp. 107--138. Springer (2008)

\bibitem{CFK+13b}
Chen, T., Forejt, V., Kwiatkowska, M., Parker, D., Simaitis, A.: Automatic
  verification of competitive stochastic systems. Formal Methods in System
  Design  \textbf{43}(1),  61--92 (2013)

\bibitem{CDT09}
Chen, X., Deng, X., Teng, S.H.: Settling the complexity of computing two-player
  {Nash} equilibria. J. ACM  \textbf{56}(3) (2009)

\bibitem{DGP09}
Daskalakis, C., Goldberg, P., Papadimitriou, C.: The complexity of computing a
  {Nash} equilibrium. Communications of the ACM  \textbf{52}(2),  89--97 (2009)

\bibitem{Z3}
De~Moura, L., Bj{\o}rner, N.: Z3: An efficient {SMT} solver. In: Proc.\
  TACAS'08. LNCS, vol.~4963, pp. 337--340. Springer (2008),
  \href{https://github.com/Z3Prover/z3}{github.com/Z3Prover/z3}

\bibitem{Dut14}
Dutertre, B.: Yices 2.2. In: Biere, A., Bloem, R. (eds.) Proc\ {CAV}'14. LNCS,
  vol.~8559, pp. 737--744. Springer (2014),
  \href{http://yices.csl.sri.com}{yices.csl.sri.com}

\bibitem{GZ89}
Gilboa, I., Zemel, E.: Nash and correlated equilibria: Some complexity
  considerations. Games and Economic Behavior  \textbf{1}(1),  80--93 (1989)

\bibitem{GNS90}
Griva, I., Nash, S., Sofer, A.: Linear and Nonlinear Optimization: Second
  Edition. CUP (01 2009)

\bibitem{gurobi}
{Gurobi Optimization, LLC}: {Gurobi Optimizer Reference Manual} (2021),
  \href{https://www.gurobi.com}{gurobi.com}

\bibitem{GHW14}
Gutierrez, J., Harrenstein, P., Wooldridge, M.J.: Reasoning about equilibria in
  game-like concurrent systems. In: Proc. 14th International Conference on
  Principles of Knowledge Representation and Reasoning (KR'14) (2014)

\bibitem{HHCN19}
Hauser, O., Hilbe, C., Chatterjee, K., Nowak, M.: Social dilemmas among
  unequals. Nature  \textbf{572},  524–527 (2019)

\bibitem{KSK76}
Kemeny, J., Snell, J., Knapp, A.: Denumerable {M}arkov Chains. Springer (1976)

\bibitem{KNPS20b}
Kwiatkowska, M., Norman, G., Parker, D., Santos, G.: Multi-player equilibria
  verification for concurrent stochastic games. In: Gribaudo, M., Jansen, D.,
  Remke, A. (eds.) Proc.\ QEST'20. LNCS, Springer (2020)

\bibitem{KNPS20}
Kwiatkowska, M., Norman, G., Parker, D., Santos, G.: {PRISM}-games 3.0:
  Stochastic game verification with concurrency, equilibria and time. In: Proc.
  CAV'20. pp. 475--487. LNCS, Springer (2020)

\bibitem{confversion}
Kwiatkowska, M., Norman, G., Parker, D., Santos, G.: Correlated equilibria and
  fairness in concurrent stochastic games. In: Fisman, D., Rosu, G. (eds.)
  Proc.\ TACAS'22. LNCS, Springer (2022)

\bibitem{KNPS21}
Kwiatkowska, M., Norman, G., Parker, D., Santos, G.: Automatic verification of
  concurrent stochastic systems. Formal Methods in System Design pp. 1--63
  (2021)

\bibitem{LRTZ06}
Littman, M., Ravi, N., Talwar, A., Zinkevich, M.: An efficient
  optimal-equilibrium algorithm for two-player game trees. In: Proc.\ {UAI}'06.
  pp. 298--305. AUAI Press (2006)

\bibitem{MM07}
McIver, A., Morgan, C.: Results on the quantitative mu-calculus {qMu}. ACM
  Trans.\ Computational Logic  \textbf{8}(1) (2007)

\bibitem{NMK+44}
von Neumann, J., Morgenstern, O., Kuhn, H., Rubinstein, A.: Theory of Games and
  Economic Behavior. Princeton University Press (1944)

\bibitem{NRTV07}
Nisan, N., Roughgarden, T., Tardos, E., Vazirani, V.: Algorithmic Game Theory.
  CUP (2007)

\bibitem{NWSL04}
Nudelman, E., Wortman, J., Shoham, Y., Leyton-Brown, K.: Run the {GAMUT}: A
  comprehensive approach to evaluating game-theoretic algorithms. In: Proc.\
  {AAMAS}'04. pp. 880--887. ACM (2004),
  \href{http://gamut.stanford.edu}{gamut.stanford.edu}

\bibitem{OR04}
Osborne, M., Rubinstein, A.: An Introduction to Game Theory. OUP (2004)

\bibitem{PNS04}
Porter, R., Nudelman, E., Shoham, Y.: Simple search methods for finding a
  {N}ash equilibrium. In: Proc.\ AAAI'04. pp. 664--669. AAAI Press (2004)

\bibitem{EP14}
Prisner, E.: Game Theory Through Examples. Mathematical Association of America,
  1 edn. (2014)

\bibitem{Put94}
Puterman, M.: Markov Decision Processes: Discrete Stochastic Dynamic
  Programming. John Wiley and Sons (1994)

\bibitem{Rab93}
Rabin, M.: Incorporating fairness into game theory and economics. The American
  Economic Review  \textbf{83}(5),  1281--1302 (1993)

\bibitem{Rab97}
Rabin, M.: Fairness in repeated games. working paper 97–252, University of
  California at Berkeley (1997)

\bibitem{SW+01}
Schwalbe, U., Walker, P.: Zermelo and the early history of game theory. Games
  and Economic Behavior  \textbf{34}(1),  123--137 (2001)

\bibitem{Sha53}
Shapley, L.: Stochastic games. PNAS  \textbf{39},  1095--1100 (1953)

\bibitem{SP17}
Stevens, S., Palocsay, S.: Teaching use of binary variables in integer linear
  programs: Formulating logical conditions. INFORMS Transactions on Education
  \textbf{18}(1),  28--36 (2017)

\bibitem{Wac09}
W{\"a}chter, A.: Short tutorial: Getting started with {\sc {i}popt} in 90
  minutes. In: Combinatorial Scientific Computing. No. 09061 in Dagstuhl
  Seminar Proceedings, Leibniz-Zentrum f{\"u}r Informatik (2009),
  \href{https://github.com/coin-or/Ipopt}{github.com/coin-or/Ipopt}

\bibitem{WB06}
W{\"a}chter, A., Biegler, L.: On the implementation of an interior-point filter
  line-search algorithm for large-scale nonlinear programming. Mathematical
  Programming  \textbf{106}(1),  25--57 (2006)

\bibitem{www}
Supporting material,
  \href{http://www.prismmodelchecker.org/files/tacas22equ/}{prismmodelchecker.org/files/tacas22equ/}

\end{thebibliography}
